\pdfoutput=1
 \documentclass[twocolumn,letterpaper]{igs-noname}

 \usepackage{igsnatbib}
  \usepackage{stfloats}

  \ifx\pdftexversion\undefined
    \usepackage[dvips]{graphicx}
  \else
    \usepackage[pdftex]{graphicx}
  \fi

\usepackage{lineno}
\usepackage{amsmath}

\begin{document}

%
%

\title[Acoustic waves in snow]{A porosity-based Biot model for acoustic waves in snow}

%
%

\author[Sidler]{Rolf Sidler}

\affiliation{%
Department of Earth Sciences, Simon Fraser University, 8888 University Drive, BC V5A1S6 Burnaby,
Canada}

%
%

\abstract{
Phase velocities and attenuation in snow can not be explained by the widely used elastic or viscoelastic models for acoustic wave propagation. Instead, Biot's model of wave propagation in porous materials should be used. However, the application of Biot's model is complicated by the large property space of the underlying porous material. Here the properties of ice and air as well as empirical relationships are used to define the properties of snow as a function of porosity. Based on these relations, phase velocities and plane wave attenuation of shear- and compressional-waves as functions of porosity or density are predicted. For light snow the peculiarity was found that the velocity of the first compressional wave is lower than the second compressional wave that is commonly referred to as the ``slow'' wave. The reversal of the velocities comes with an increase of attenuation for the first compressional wave. This is in line with the common observation that sound is strongly absorbed in light snow.
The results have important implications for the use of acoustic waves to evaluate snow properties and to numerically simulate wave propagation in snow.
}

%
%

\maketitle

%
%

\section{Introduction}

The use of acoustics to investigate snow is complicated by the fact that sometimes lower wave velocities can be observed with increasing density of the snow \citep{oura:1952}. This observation is at odds with elastic or visco-elastic wave propagation theory, for which higher velocities are expected for the considerably higher bulk and shear moduli of denser snow.
Yet, the observed wave velocities can be explained with wave propagation theory for porous materials, where a second compressional wave, also known as ``slow'' wave, is predicted \citep{johnson:1982,smeulders:2005}.

\citet{oura:1952, smith:1969, yamada:1974} measured acoustic wave velocities and attenuation in field and laboratory environments.
\citet{gubler:1977} measured acceleration in the snowpack and air pressure above the snowpack for explosives used in avalanche mitigation operations.
\citet{johnson:1982} successfully used Biot's model for wave propagation in porous materials to predict wave velocities in snow.
\citet{sommerfeld:1983} observed increased acoustic emissions from unstable snowpacks compared to acoustic emissions of stable snowpacks.
\citet{mellor:1975} and \citet{shapiro:1997} published extensive reviews on snow mechanics including acoustic wave propagation and proposed wave velocity as a potential index property for snow.
Amongst others, \citet{buser:1986,attenborough:1988,marco:1996,marco:1998,maysenholder:2012} investigated acoustic impedance and attenuation of snow based on the  so called ``rigid-frame'' model \citep{terzaghi:1923,zwikker:1947} in which the wave traveling in the pore space is completely decoupled from the wave traveling in the frame of the porous material.
Recently acoustic methods have been used to monitor and spatially locate avalanches \citep{surinach:2000, herwijnen:2011,lacroix:2012}, to estimate the height and sound absorption of snow covering ground \citep{albert:2001,albert:2009,albert:2013} and to estimate the snow water equivalent of dry snowpacks \citep{kinar:2009}. 
\citet{kapil:2014} used metallic waveguides to measure acoustic emissions from deforming snowpacks.
 
The advantage of the rigid frame model is that it is relatively straight forward to extract tortuosity of the pore space and pore fluid properties from the phase velocities of the slow wave.
Applications are widespread and range from non-destructive testing, medical applications and soil characterization to sound absorption \citep{fellah:2004,jocker:2009,shin:2013,attenborough:2013}.

The rigid-frame model can be deduced from Biot's \citeyearpar{biot:1956,biot:1962} theory under the assumption that the stiffness of the porous frame is considerably higher than the stiffness of the pore fluid.
Consequently, the rigid-frame model does not account for the interaction between the pore fluid and the porous frame as does Biot's theory. Also the viscous effects of the pore fluid are approximated with complex moduli in the rigid-frame model and consequently accounted for with a phenomenological instead of a physical model as in Biot's theory where the viscous friction of the fluid moving relative to the solid frame is causing the observed attenuation.
Especially in light snow and in wet snow, where the frame and the stiffness of the pore fluid are of comparable order of magnitude Biot theory is expected to provide superior results than the rigid-frame model \citep{hoffman:2012}. 
Also a physical model is to be preferred over a phenomenological as the results can be compared to complementary measurements and consequently has more predictive power. 
However, a disadvantage for the application of Biot's model is the large number of properties that have to be specified.

While phase velocities obtained with plane wave solutions for Biot's theory tend to correspond relatively well with its measured counterparts, the plane wave attenuation can generally not be readily compared.
The wave attenuation is complicated by the superposition of effects that all lead to a decrease of wave amplitude and are difficult to separate.
Plane wave attenuation does, for example, not account for geometrical spreading, that strongly depends on the  geometry of the experiment and is present in virtually all physical measurements. 

Here, we propose a porous snow model as a function of porosity and use it to estimate the wave velocities and  attenuation of compressional and shear wave modes using plane wave solutions for Biot's \citeyearpar{biot:1956} differential equation of wave propagation in porous materials.  We compare the results to measurements from the literature and  investigate the sensitivity of the fast and slow compressional waves to individual parameters of the porous model such as, for example, specific surface area (SSA).

\section{Methods}

\subsection{Porous material properties for snow}
\label{sec:snow-model}

An inherent problem when working with Biot-type porous models is the large number of material properties involved. 
To address this problem empirical relationships and {\it a priori} information is gathered in this section to express the porous material properties as a function of porosity.

A Biot-type porous material is characterized by ten properties with porosity arguably being the most significant. For the stress strain relations the fluid bulk modulus $K_f$, the bulk modulus of the frame material $K_s$, the bulk modulus of the matrix $K_m$, and the shear modulus $\mu_s$ have to be known.
The equations of motion require the densities of the solid and fluid materials $\rho_s$ and $\rho_f$, the porosity $\phi$, and the tortuosity $\mathcal{T}$.
The energy dissipation due to the relative motion of the fluid to the solid are based on Darcy's \citeyearpar{darcy:1856} law and requires the knowledge of the permeability $\kappa$ and the viscosity $\eta$ of the pore fluid.  

Typical values for Young's modulus of ice, the frame material of snow, are between 9.0~GPa and 9.5~GPa with a Poisson's ratio of $\pm$ 0.3 \citep{hobbs:1974,mellor:1983,schulson:1999}.
The Young's modulus $E$ can be converted to bulk modulus $K$ as
\citep{mavko:2009}
\begin{linenomath*}
\begin{equation}
\label{eq:bulk}
K = \frac{E}{3(1-2\nu)} ,
\end{equation}
\end{linenomath*}
where $\nu$ is the Poisson's ratio.
The resulting frame bulk modulus $K_s$ for snow is approximately 10~GPa.

\begin{figure}
\centering{\includegraphics[width=80mm]{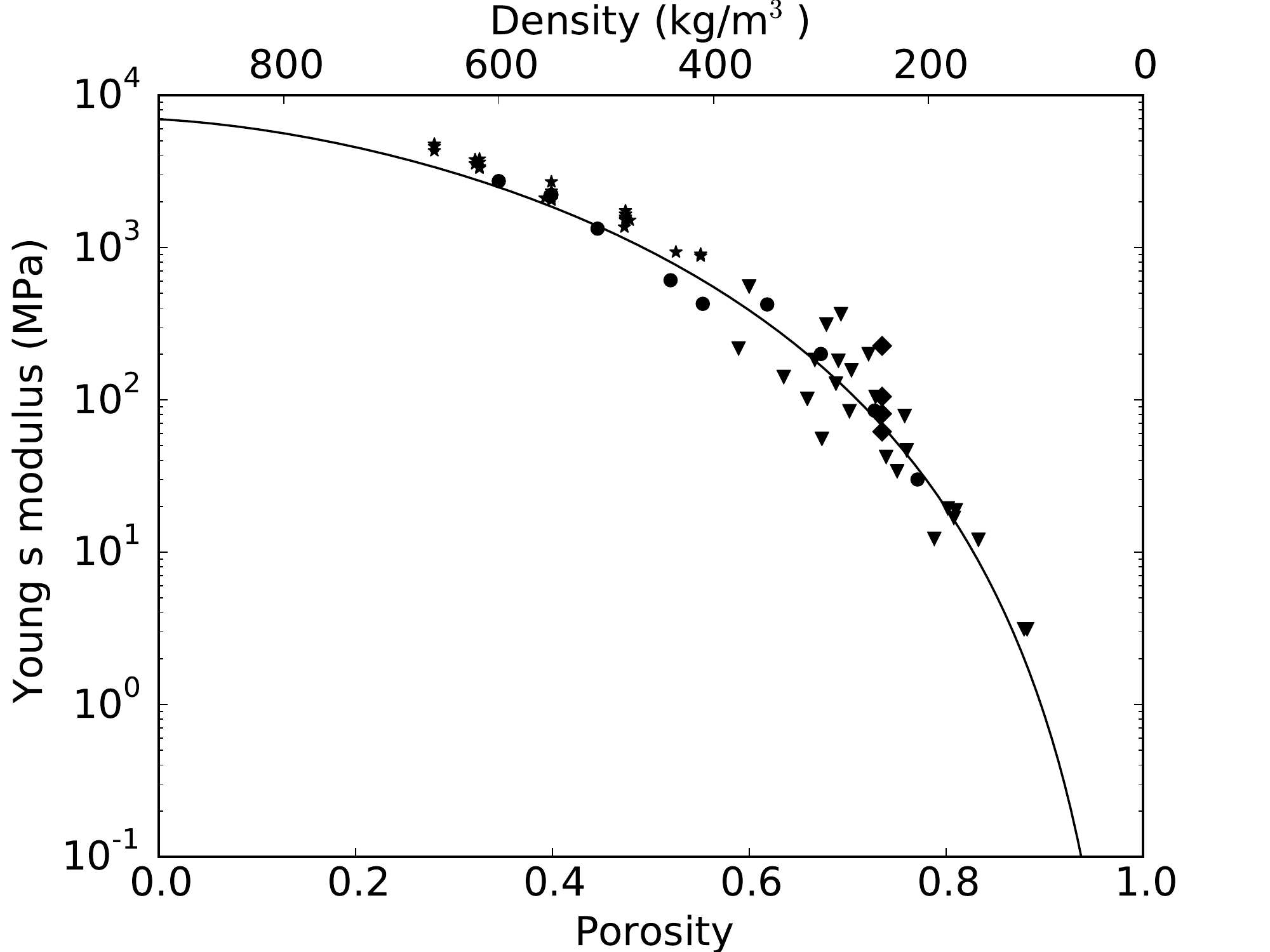}}
\caption{  
Krief equation fitted to dynamic measurements (solid line) and compared to dynamic measurements from \citet{johnson:1982} and \citet{smith:1969} indicated with circles and stars, respectively. Theoretical values obtained from numerical modeling of microtomography snow structures are indicated with diamonds and triangles for \citet{schneebeli:2004} and \citet{reuter:2013}, respectively.}
\label{fig:bulkmodulus}
\end{figure}

For the snow matrix bulk modulus $K_m$, the Krief equation \citep{garat:1990,mavko:1998} 
\begin{linenomath*}
\begin{equation}
\label{eq:krief}
K_{\rm m} = K_{\rm s} (1-\phi)^\frac{4}{(1-\phi)} ,
\end{equation}
\end{linenomath*}
can be parameterized as
\begin{linenomath*}
\begin{equation}
\label{eq:modkrief}
K_{\rm m} = K_{\rm s} (1-\phi)^\frac{a}{(b-\phi)},
\end{equation}
\end{linenomath*}
and values for $a=30.85$ and $b=7.76$ can be obtained by a least square inversion on the measurements presented by \citet{johnson:1982}.
In Figure \ref{fig:bulkmodulus} the Young's moduli resulting from Equation (\ref{eq:modkrief}) are shown in comparison with measurements and theoretical estimates of Young's moduli \citep{johnson:1982,smith:1969,schneebeli:2004,reuter:2013}.

\begin{figure}
\centering{\includegraphics[width=80mm]{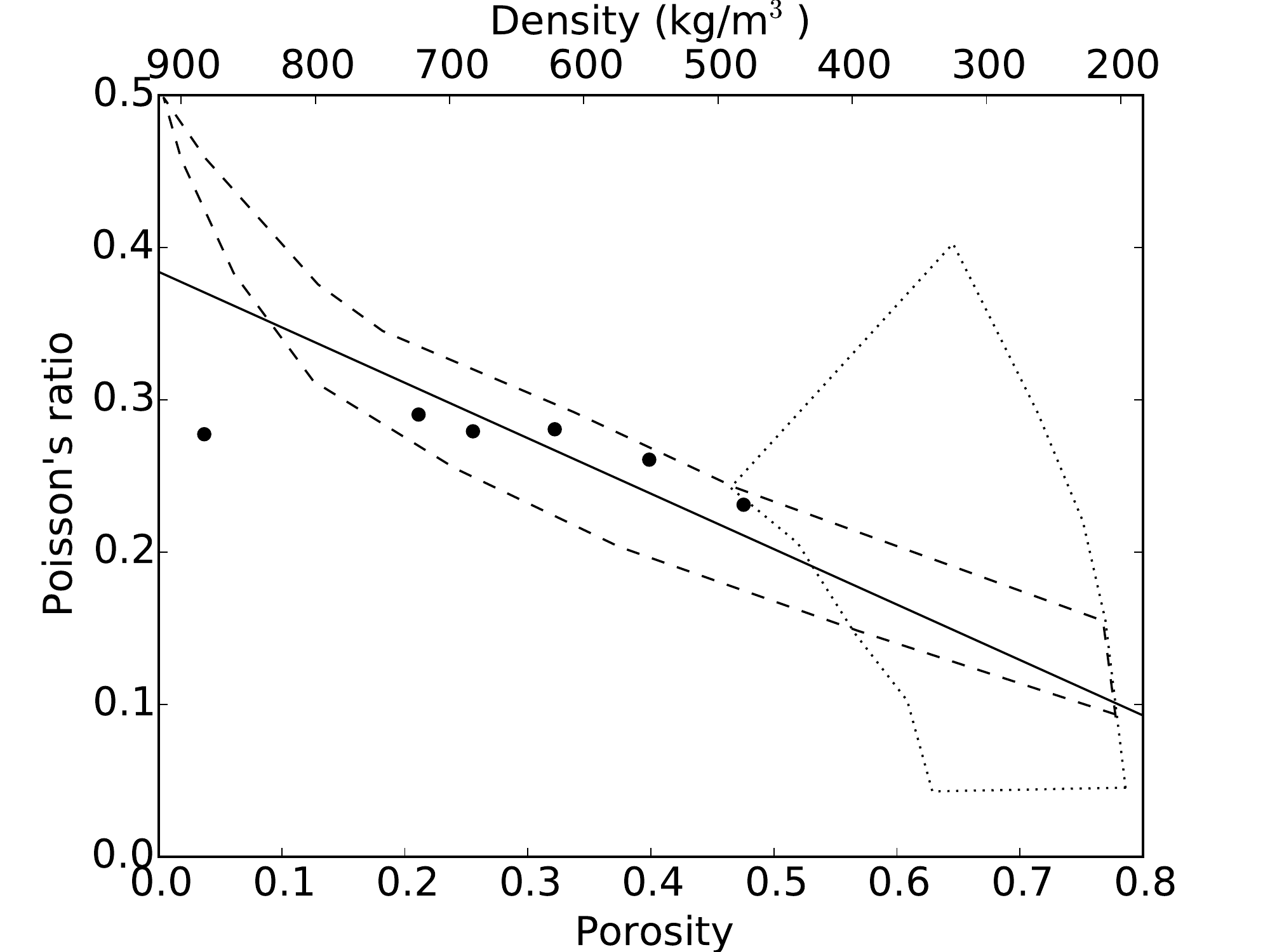}}
\caption{ 
Poisson's ratio as a function of porosity according to Equation (\ref{eq:poisson}) compared to measurements from \citet{bader:1952} (dashed lines), \citet{roch:1948} (dotted lines), and \citet{smith:1969} (circles).}
\label{fig:poisson-ratio}
\end{figure}

\begin{figure}
\centering{\includegraphics[width=80mm]{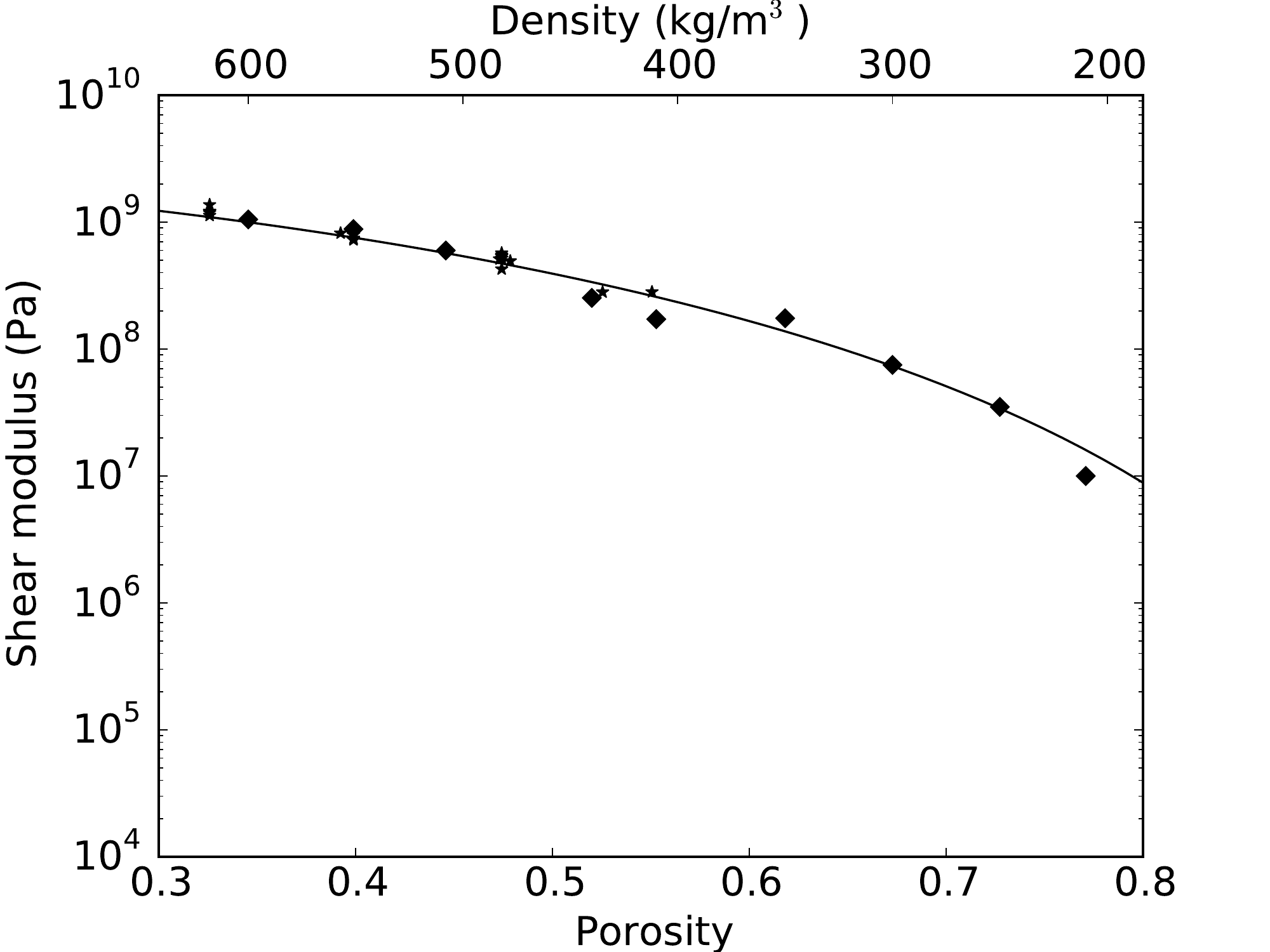}}
\caption{
Shear moduli for snow as a function of porosity by using Equations (\ref{eq:modkrief}) and (\ref{eq:poisson})  compared to measurements presented by \citet{johnson:1982} (diamonds) and \citet{smith:1969} (stars).}
\label{fig:shear-modulus}
\end{figure}

In combination with the Poisson's ratio of snow, Equation (\ref{eq:modkrief}) can also be used to estimate shear moduli of snow as a function of porosity by using the relationship \citep{mavko:2009}
\begin{linenomath*}
\begin{equation}
\label{eq:shear}
\mu_s = \frac{3}{2} \frac{K_m (1-2 \nu)}{1+\nu} .
\end{equation}
\end{linenomath*}
For this purpose the linear relationship
\begin{linenomath*}
\begin{equation}
\label{eq:poisson}
\nu = 0.38 - 0.36 \, \phi , 
\end{equation}
\end{linenomath*}
is used to express the Poisson's ratio of snow as a function of porosity. Figure \ref{fig:poisson-ratio} shows how this function relates to measurements from \citet{bader:1952}, \citet{roch:1948}, and \citet{smith:1969}.
%
%
The shear moduli resulting from Equations (\ref{eq:modkrief}), (\ref{eq:shear}) and, (\ref{eq:poisson}) are shown in Figure \ref{fig:shear-modulus} and compared to measurements from \citet{johnson:1982} and \citet{smith:1969}.

The tortuosity ${\cal T}$ describing the `twisting' of the actual flow path of the pore fluid compared to a straight line can be estimated based on geometrical considerations as
\begin{linenomath*}
\begin{equation}
\label{eq:berrymann}
{\cal T} = 1- s \left( 1-\frac{1}{\phi} \right) ,
\end{equation}
\end{linenomath*}
where $s$ is the so called shape factor \citep{berryman:1980}.
For a packing of sphere the shape factor is 0.5 and Equation (\ref{eq:berrymann}) reduces to 
\begin{linenomath*}
\begin{equation}
{\cal T} = \frac{1}{2} \left( 1+\frac{1}{\phi} \right).
\end{equation}
\end{linenomath*}
%

%

The permeability is estimated using the Kozeny-Carman relation 
\begin{linenomath*}
\begin{equation}
\label{eq:CK}
\kappa = C \frac{r^2 \phi^3}{(1-\phi)^2} ,
\end{equation}
\end{linenomath*}
where  $\kappa$ is the permeability, $\phi$ the porosity, and $C$ is an empirical constant \citep{mavko:1998}. For sediments the constant is $C_{\rm s}=0.003$ \citep{mavko:1997,carcione:2006a}, but one order of magnitude larger for snow $C_{\rm B}=0.022$ \citep{calonne:2012,bear:1972}.
The grain diameter $r$ can be related to the specific surface area (SSA) of snow with the relation 
\begin{linenomath*}
\begin{equation}
\label{eq:avgrain}
r = \frac{3}{ \text{SSA} \rho_{i}},
\end{equation}
\end{linenomath*}
where $\rho_{i}$ is the density of ice.
Substituting Equation (\ref{eq:avgrain}) into Equation (\ref{eq:CK}) leads to
\begin{linenomath*}
\begin{equation}
\label{eq:KC-red}
\kappa = 0.2 \frac{\phi^3}{(\text{SSA})^2 (1-\phi)^2} .
\end{equation}
\end{linenomath*}

Due to the compaction and metamorphosis processes inherent to snow it can be assumed that the specific surface area by itself is a function of porosity \citep{legagneux:2002,herbert:2005a}.  
\citet{domine:2007} use the equation
\begin{linenomath*}
\begin{equation}
\label{eq:SSA}
SSA =  -308.2 \ln(\rho_S) -206.0 ,
\end{equation}
\end{linenomath*}
to relate specific surface area to snow density.
For dry snow the density $\rho_s$ can be obtained from snow with porosity $\phi$ as
\begin{linenomath*}
\begin{equation}
\rho_s = (1-\phi) \cdot 916.7 \ \frac{\text{kg}}{\text{m$^3$}} .
\end{equation}
\end{linenomath*}
The equation (\ref{eq:SSA}) yields negative values for the specific surface area for porosities smaller than $\phi = 0.44$.
Therefore the relationship is used here only for porosities larger than $\phi = 0.65$ and a constant value for specific surface area of 15~m$^{2}$/kg is used for lower porosities. 
In this study Equation (\ref{eq:SSA}) is intended to reflect an average trend and is shown together with high and low constant values for specific surface area to illustrate the possible variability.

The density $\rho_f$, viscosity $\eta$, and bulk modulus $K_f$ of air as the pore fluid of snow are assumed to be constant, which means independent of temperature and altitude and are given in Table \ref{tbl:parameters} \citep{lide:2005}. 

\begin{table}
\internallinenumbers
\caption{Pore fluid properties of air \citep{lide:2005}.}
\label{tbl:parameters}
\centering
\begin{tabular}{ll}
\hline 
 density, $\rho_f$ & 1.30 kg/m$^{3}$ \\
 viscosity, $\eta$ & $1.7 \cdot 10^{-5}$ Pa s \\
 bulk modulus, $K_f$ & $1.42 \cdot 10^{5}$ Pa \\
\hline
\end{tabular}
\end{table}

\subsection{Phase velocities and plane wave attenuation}
\label{sec:planewave}

A convenient way to get closed form solutions for Biot's \citeyearpar{biot:1956a} differential equations is to assume plane waves solutions and substitute those into the differential equations. The complex plane wave modulus is then obtained by solving the resulting dispersion relation \citep{johnson:1982,pride:2005,carcione:2007}.
As in the poroelastic case the dispersion relation is a quadratic equation, there are two roots that correspond to the first and second compressional waves.
The phase velocity $V$ and attenuation in terms of the dimensionless quality factor $Q$ can then be obtained from the complex plane wave modulus $V_c$ as \citep{oconnel:1978}
\begin{linenomath*}
\begin{equation}
\label{phasevel}
V (\omega) = \left[ \mbox{Re}(V_c (\omega)^{-1} )  \right]^{-1} ,
\end{equation}
\end{linenomath*}

\begin{linenomath*}
\begin{equation}
\label{attenuattion}
Q_p (\omega)^{-1} = 2 \frac{\mbox{Im}(V_c (\omega) )}{\mbox{Re}(V_c (\omega) )} ,
\end{equation}
\end{linenomath*}
where $\omega$ is the angular frequency.

The solutions are generally computed for individual frequencies and porosities. 
However, under the assumption that any dissipative effects are ignored and that the bulk modulus of the fluid is much smaller than the bulk modulus of the solid matrix it can be shown that the first compressional wave velocity $V_{\infty1}$ can be expressed as
\begin{linenomath*}
\begin{equation}
\label{eq:fast-wave}
V_{\infty1} = \sqrt{\frac{E_m}{\rho - \phi \rho_f / \cal{T}}},
\end{equation}
\end{linenomath*}
and the second compressional wave velocity $V_{\infty2}$ as
\begin{linenomath*}
\begin{equation}
\label{eq:slow-wave}
V_{\infty2} = \sqrt{\frac{K_f}{\rho_f \cal{T}}},
\end{equation}
\end{linenomath*}
where $E_m = K_m + (4/3) \mu_s$ \citep[][p. 81]{bourbie:1987}.
While these expressions may deviate from the Biot phase velocities, they illustrate that the first compressional wave travels mainly in the skeleton and the second compressional wave mainly in the fluid. 
The first compressional wave is therefore also most sensitive to the properties of the ice matrix. The second compressional wave is sensitive mostly to properties of the pore space and the pore fluid.

\subsection{Dynamic viscous effects}
\label{sec:viscous}

The fluid flow in the pores of the material has a different character for lower and higher frequencies \citep{biot:1956a}. For lower frequencies the flow is of Poisson-type where the flow is fastest in the center of a pore and reduces gradually in a parabolic shape towards the outside of the pores. For higher frequencies the importance of inertial forces increases. The fluid in the center of the pores flows all with the same velocity like an ideal fluid, while the fluid at the outside of the pores remains attached to the pore walls. In between the two forms the so called viscous boundary layer \citep{pride:2005}.
The transition between the low and high frequency flow behavior occurs when the viscous boundary layers are smaller than the pore diameter. The frequency at which this transition occurs is called the Biot frequency $f_{Biot}$ and can be computed as
\begin{linenomath*}
\begin{equation}
\label{eq:biotfrequency}
f_{Biot} = \frac{\eta \phi}{2 \pi {\cal T} \kappa \rho_f},
\end{equation}
\end{linenomath*}
where $\eta$ and $\rho_f$ are the viscosity and density of the pore fluid, and $\cal{T}$ and $\kappa$ are the tortuosity and the permeability of the pore space, respectively \citep[][p. 270]{carcione:2007}.

\begin{figure}
\centering{\includegraphics[width=80mm]{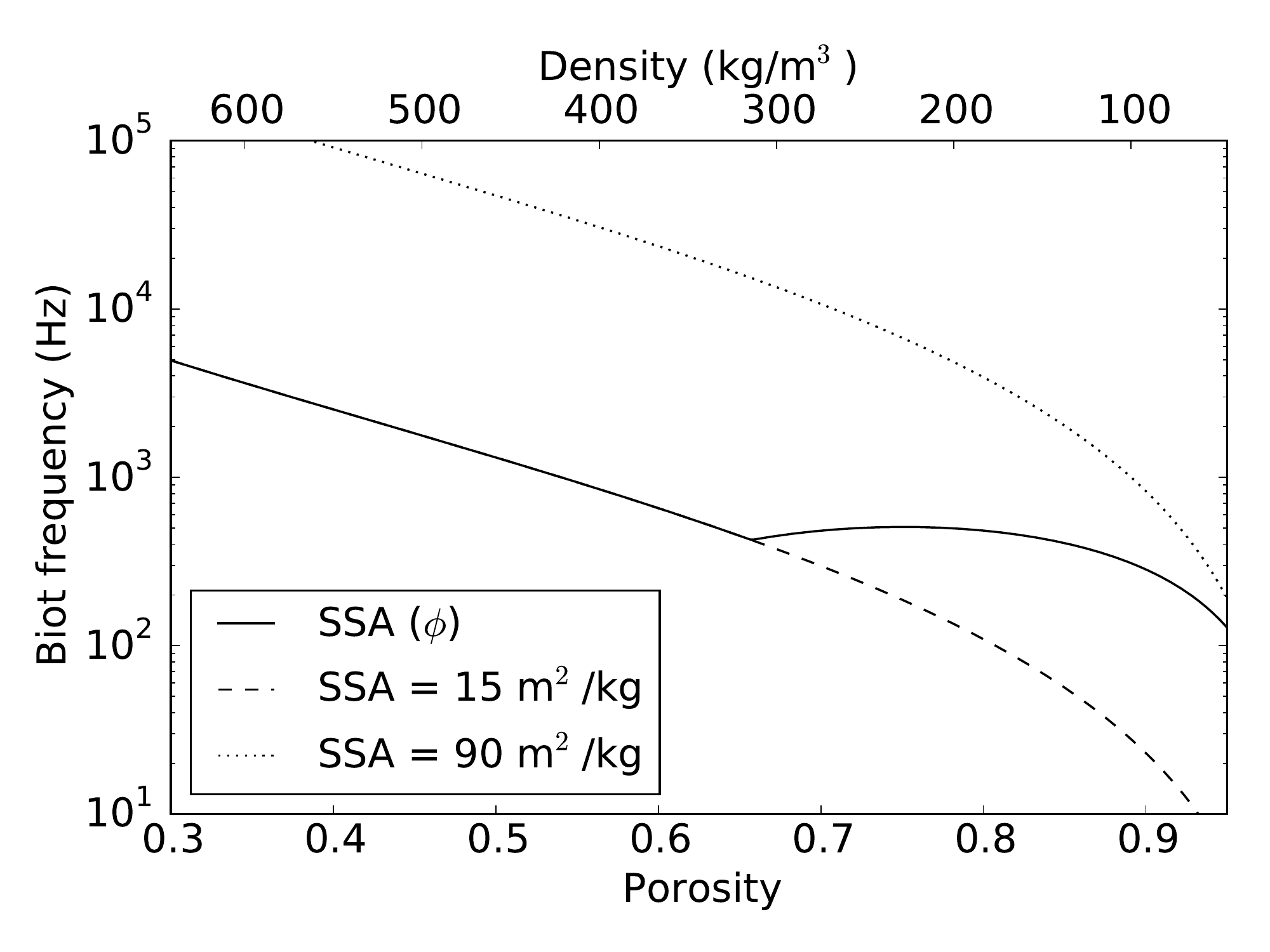}}
\caption{Biot's characteristic frequency as a function of porosity based on the relations presented in Section \ref{sec:snow-model}. The solid line corresponds to frequency dependent variation of specific surface area according to Equation \ref{eq:SSA} and the dashed and dotted line corresponds to constant values of SSA = 15 m$^{2}$/kg and SSA = 90 m$^{2}$/kg, respectively}
\label{fig:biotfrequency}
\end{figure}

Biot's \citeyearpar{biot:1956} theory is considered valid for frequencies up to the Biot frequency. For higher frequencies \citet{johnson:1987} introduced a frequency dependent permeability that accounts for the different flow behavior in the low and high frequency limit and is often referred to as a frequency correction or the JDK model.
Figure \ref{fig:biotfrequency} shows the Biot frequency for snow based on the relationships between porosity and the involved material properties as presented in Section \ref{sec:snow-model}, where the permeability further depends on the specific surface area. To illustrate the variability of Biot's frequency due to specific surface area the Biot frequency is plotted for Equation (\ref{eq:SSA}) which depends on porosity and for  constant values of SSA = 15 m$^{2}$/kg and SSA = 90 m$^{2}$/kg.

The results shown in this study are evaluated using the frequency correction \citep{johnson:1987}. However, for the first compressional wave there are virtually no differences when the frequency correction is neglected. For the second compressional wave the differences are rather low, except for frequencies in the range of the Biot frequency, where moderate differences can be observed.

\section{Results}

\begin{figure}
\centering{\includegraphics[width=80mm]{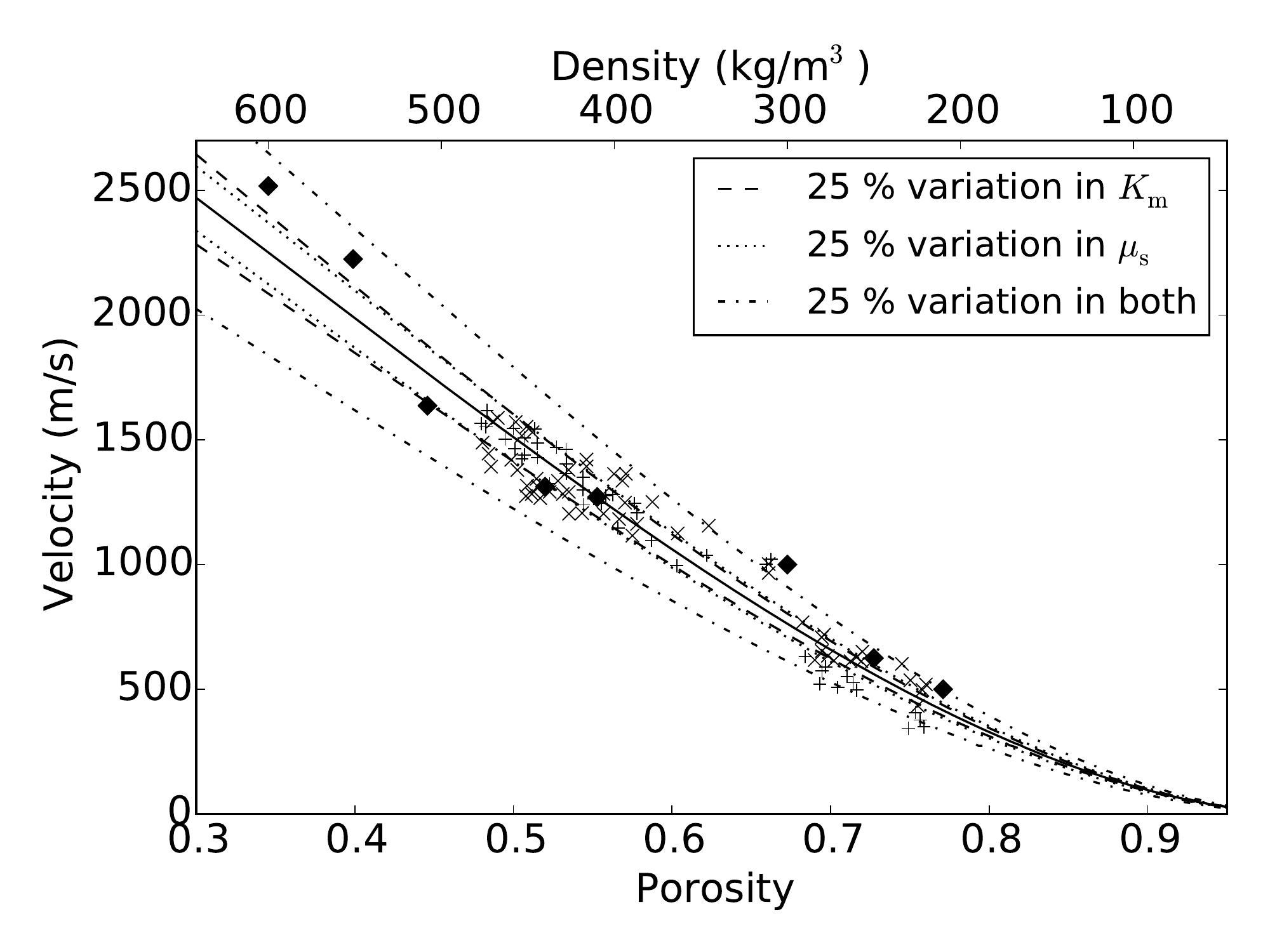} }
\caption{
Predicted phase velocities for the compressional wave of the first kind at 1 kHz compared to measurements from \citet{johnson:1982} (diamonds) and \cite{smith:1969} (crosses). The dashed and dotted lines correspond to predicted velocities for a 25~\% variation of matrix bulk modulus and shear modulus, respectively. The dash-dot line corresponds to a  25~\% variation in both.}
\label{fig:fast-velocity} 
\end{figure}

In this section phase velocities and plane wave attenuation for snow are presented as a function of porosity based on the relationships presented in Section \ref{sec:snow-model}.
Figure \ref{fig:fast-velocity} shows the predicted phase velocity for the first compressional wave and a frequency of 1~kHz as a function of porosity. The predicted phase velocities are compared to measurements from \citet{smith:1969} and \citet{johnson:1982}.
In addition, the predicted velocity for an individual and a combined variation of 25 \% in bulk and shear modulus are shown.
The velocity strongly decreases with increasing porosity and the variation of bulk and shear modulus for snow of the same porosity are small compared to the change of velocity over the porosity range.

\begin{figure}
\centering{\includegraphics[width=80mm]{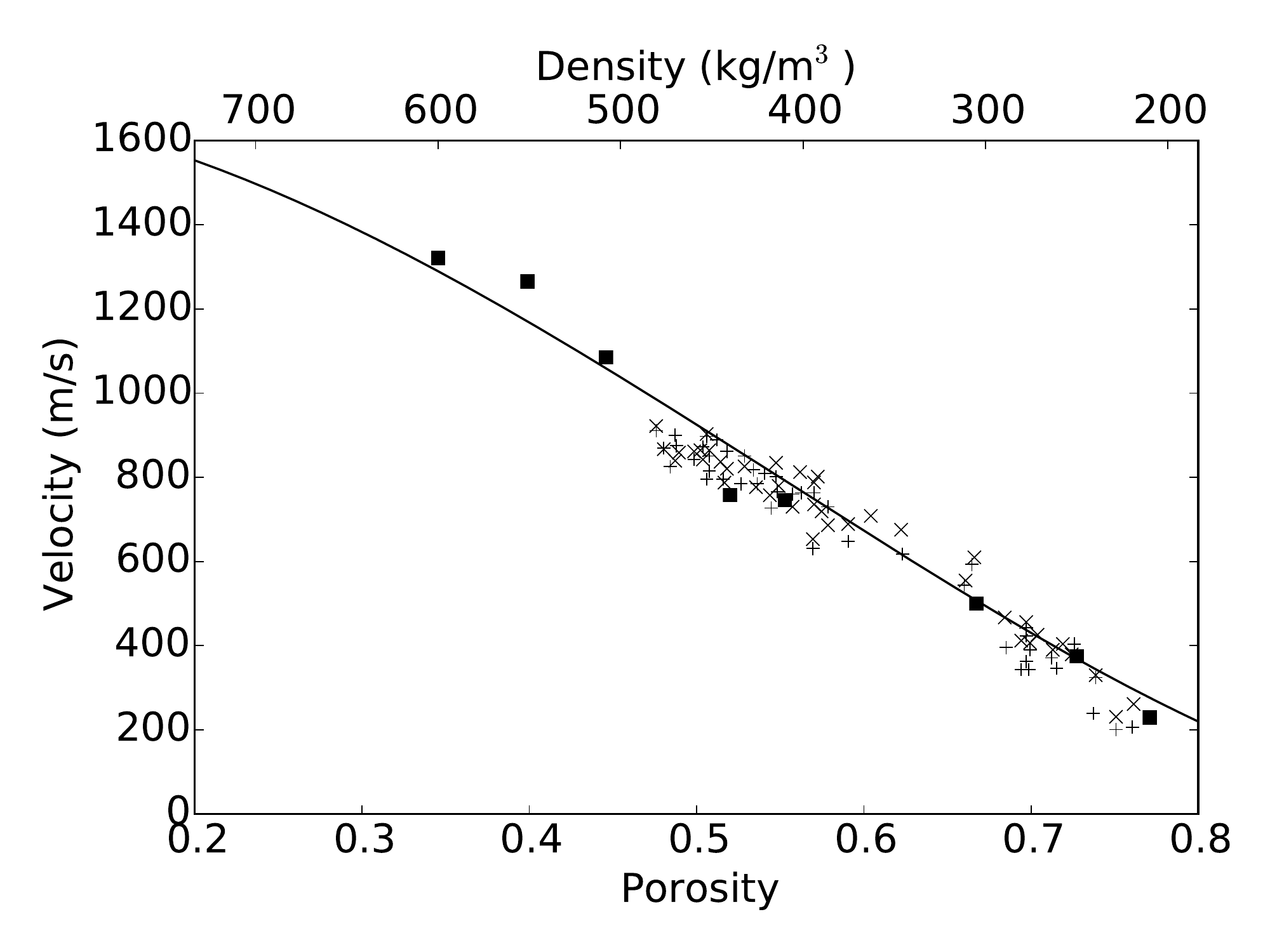}}
\caption{
Predicted shear velocities at 1kHz. Squares and crosses correspond to shear wave velocity measurements from \citet{johnson:1982} and \citet{yamada:1974}, respectively.}
\label{fig:shear-velocity}
\end{figure}

The predicted shear velocities at 1~kHz are compared to measurements by \citet{johnson:1982} and \citet{yamada:1974} in Figure \ref{fig:shear-velocity}.
Similar to the first compressional wave, the shear velocity strongly decreases with increasing porosity.

\begin{figure}
\centering{\includegraphics[width=80mm]{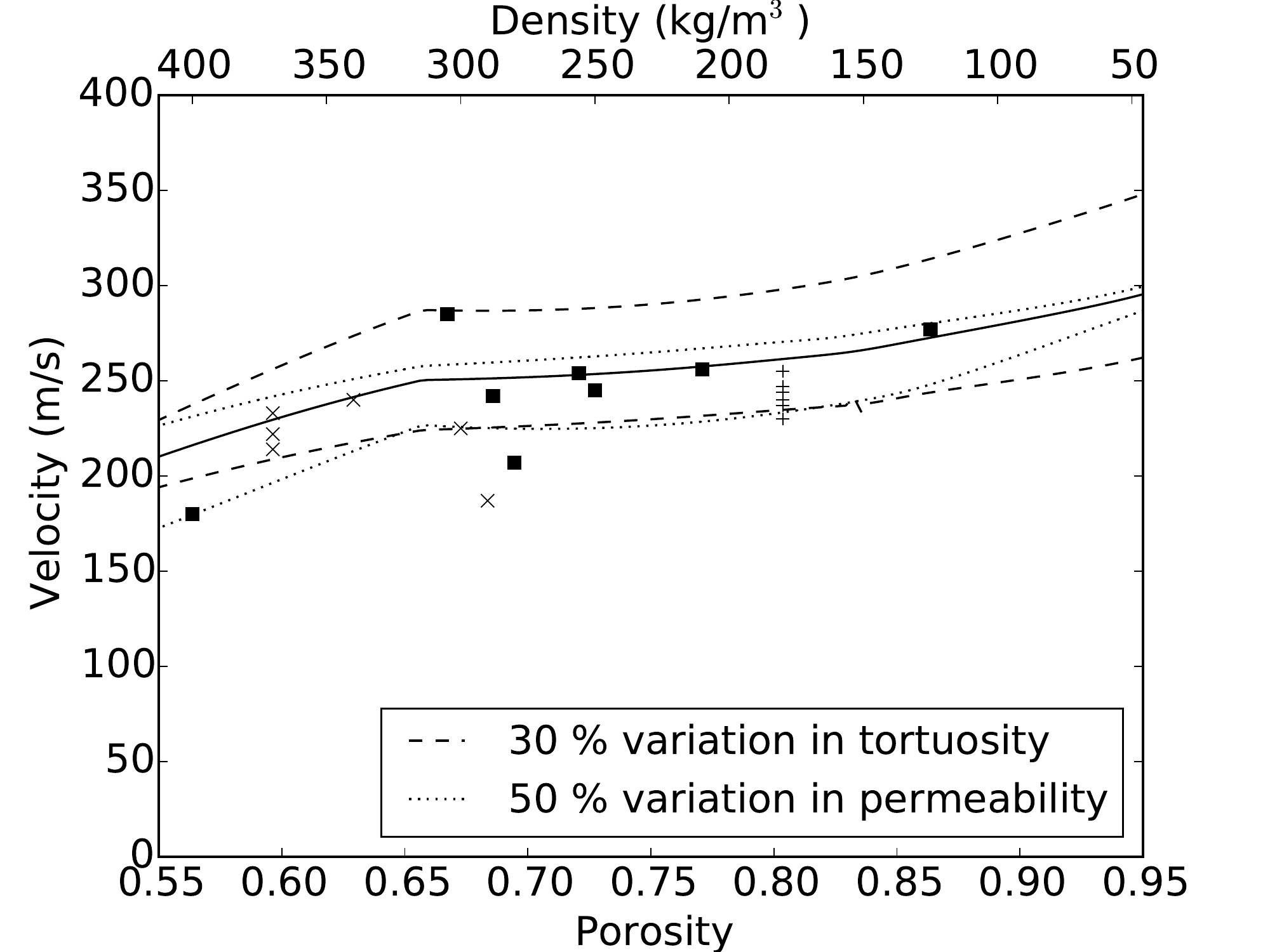}}
\caption{
Predicted phase velocities for the second compressional wave at 500 Hz. The dashed and dotted lines correspond to phase velocities for 30~\% variation in tortuosity and 50~\% variation in permeability, respectively.
Squares represent wave velocity measurements from \citet{johnson:1982}. Crosses correspond to measurements from \cite{oura:1952}.
Please note that an increasing tortuosity decreases the velocity while an increase of permeability increases the velocity of the second compressional wave.}
\label{fig:slow-sens}
\end{figure}

The predicted phase velocities of the second compressional wave at 500~Hz as a function of porosity are shown in Figure \ref{fig:slow-sens} and are compared to measurements from \citet{oura:1952} and \citet{johnson:1982}. 
As the pore fluid properties are assumed to be constant, the phase velocity of the second compressional wave depends almost exclusively on variations in permeability and tortuosity. Variation in frame bulk modulus and shear modulus have virtually no influence on phase velocity and attenuation of the second compressional wave.
The tortuosity has a stronger lever on the phase velocity than the permeability and 30 \% variation in tortuosity leads to larger changes in phase velocity than a 50 \% variation in permeability.
The phase velocity of the second compressional wave shows only little variation with porosity and is mainly sensitive to the geometrical structure of the pore space.

\begin{figure}
\centering
\begin{minipage}[t]{3mm}
\large a) 
\end{minipage}
\begin{minipage}[t]{80mm}
\vspace{-10pt}
\includegraphics[width=1\textwidth]{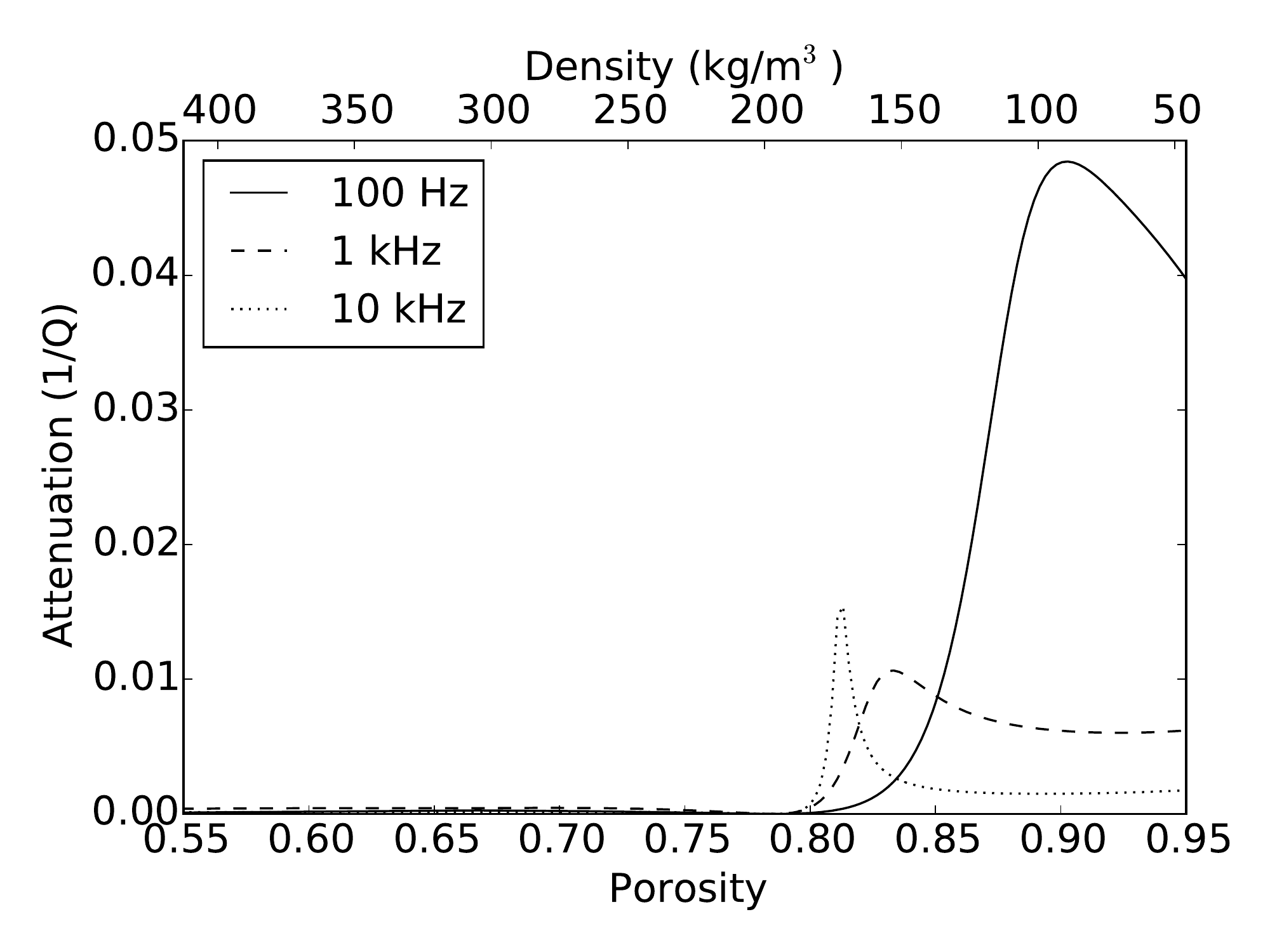}
\end{minipage}
\begin{minipage}[t]{3mm}
\large b) 
\end{minipage}
\begin{minipage}[t]{80mm}
\vspace{-10pt}
\includegraphics[width=1\textwidth]{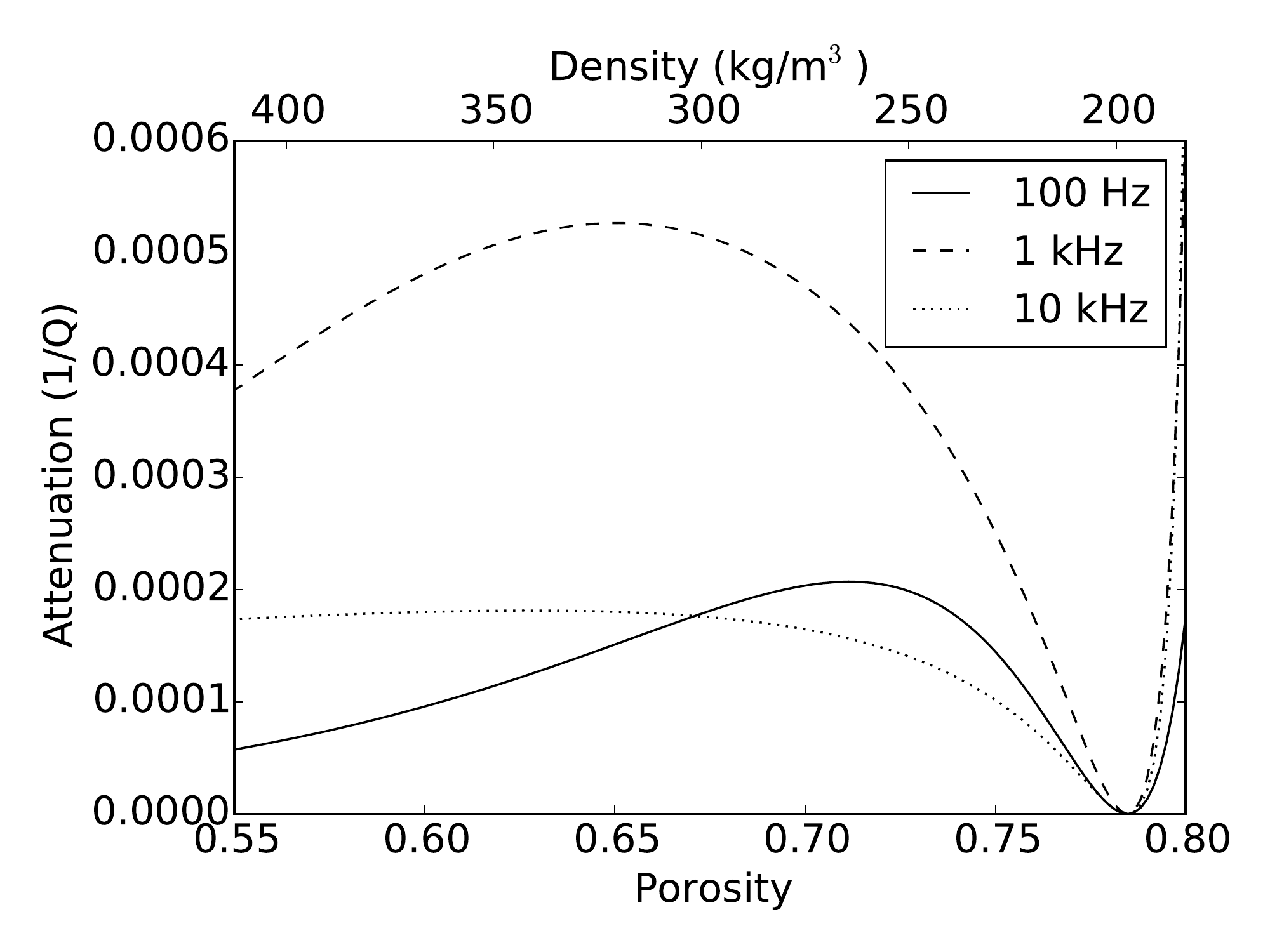}
\end{minipage} \\ 
\caption{
Predicted attenuation for the first compressional wave as a function of porosity. Figure b) shows a fragment of a) for porosities below $\phi = 0.8$. The attenuation of the first compressional wave is orders of magnitudes higher for light snow than for denser snow.}
\label{fig:fast-att} 
\end{figure}

The plane wave attenuation for the first compressional wave as a function of porosity is shown for three different frequencies in Figure \ref{fig:fast-att}. 
The plane wave attenuation of the first compressional wave is orders of magnitude higher for light snow with a porosity $\phi \gtrsim 0.8$ than for denser snow. 
Figure \ref{fig:fast-att}b) therefore shows the porosity range between $\phi$ = 0.55 and $\phi$ = 0.8 where variations in attenuation can not be resolved in Figure \ref{fig:fast-att}a).

\begin{figure}
\centering
\begin{minipage}[t]{3mm}
\large a) 
\end{minipage}
\begin{minipage}[t]{80mm}
\vspace{-10pt}
\includegraphics[width=1\textwidth]{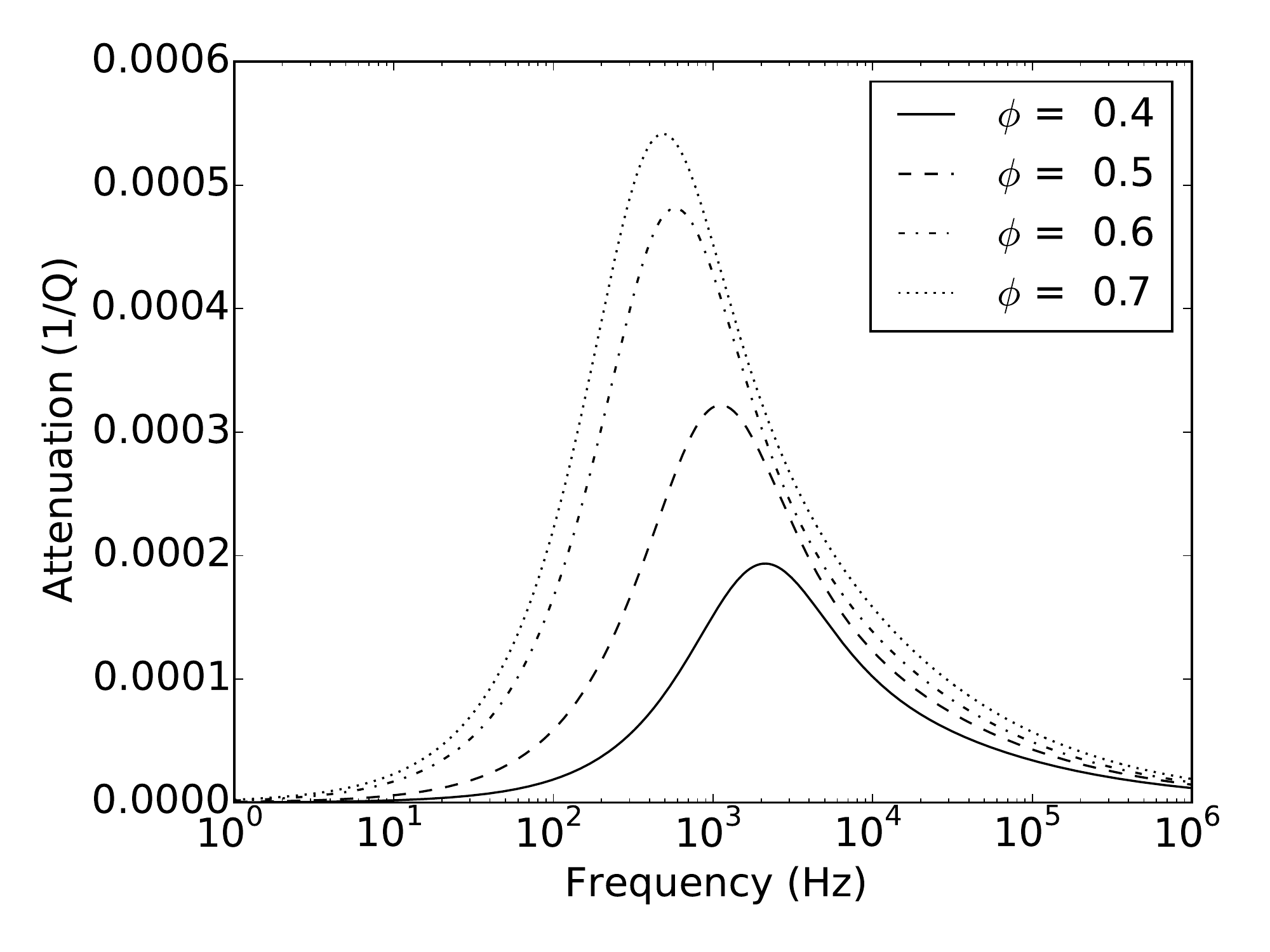}
\end{minipage}
\begin{minipage}[t]{3mm}
\large b) 
\end{minipage}
\begin{minipage}[t]{80mm}
\vspace{-10pt}
\includegraphics[width=1\textwidth]{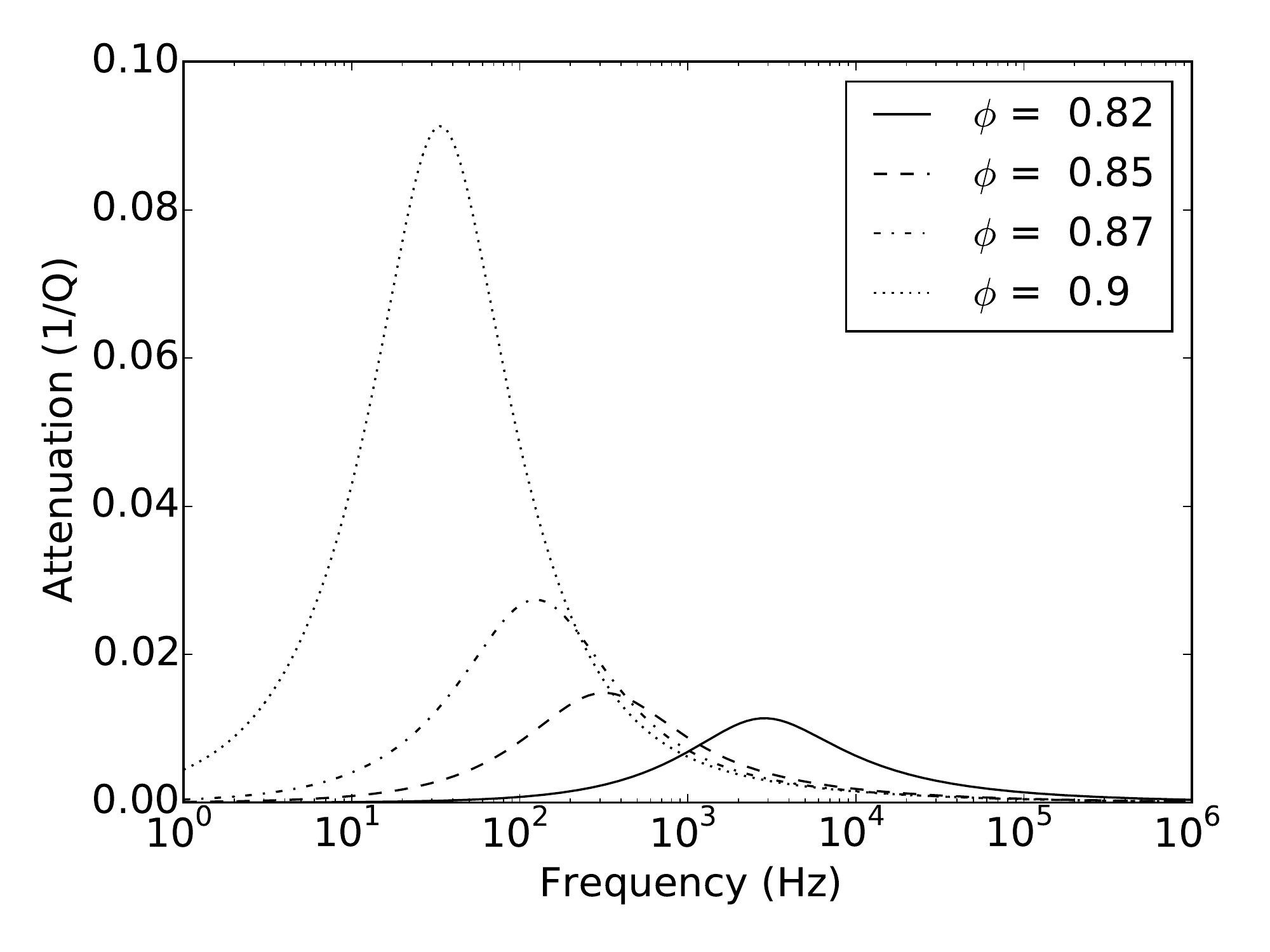}
\end{minipage} \\ 
\caption{ 
Frequency dependent attenuation for for the first compressional wave in (a) medium to dense and (b) light snow. The peak of the attenuation shifts toward higher frequencies for denser snow. Please note that the amplitude of the attenuation is orders of magnitude larger for light snow with a porosity $\phi \gtrsim 0.8$.}
\label{fig:att-peak}
\end{figure}

Homogeneous Biot-type porous materials are known to have a characteristic peak of attenuation \citep{geertsma:1961,carcione:2006a}.
In Figure \ref{fig:att-peak} these attenuation peaks are shown for snow of different densities. 
Again, the figure is split into two subfigures to account for the significant difference of attenuation levels for light and dense snow that was already apparent in Figure \ref{fig:fast-att}.
Peak attenuation shifts towards lower frequencies and the attenuation level increases with increasing porosity.
The same is true for light snow but with considerably higher attenuation levels. Also the peak attenuation frequency overlap for a porosity range around $\phi = 0.8$.

%

\begin{figure}
\centering
\begin{minipage}[t]{3mm}
\large a) 
\end{minipage}
\begin{minipage}[t]{80mm}
\vspace{-10pt}
\includegraphics[width=1\textwidth]{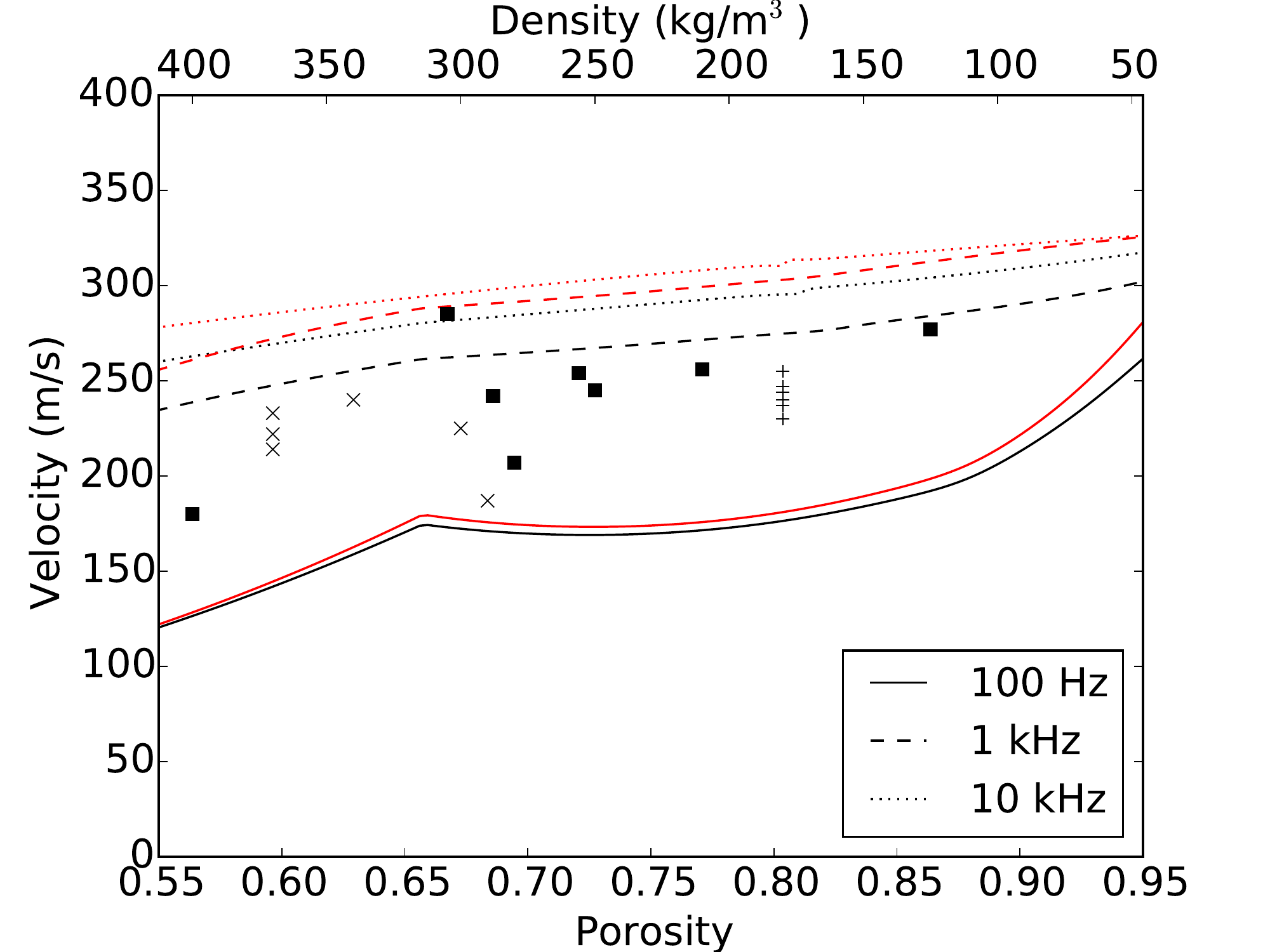}
\end{minipage}
\begin{minipage}[t]{3mm}
\large b) 
\end{minipage}
\begin{minipage}[t]{80mm}
\vspace{-10pt}
\includegraphics[width=1\textwidth]{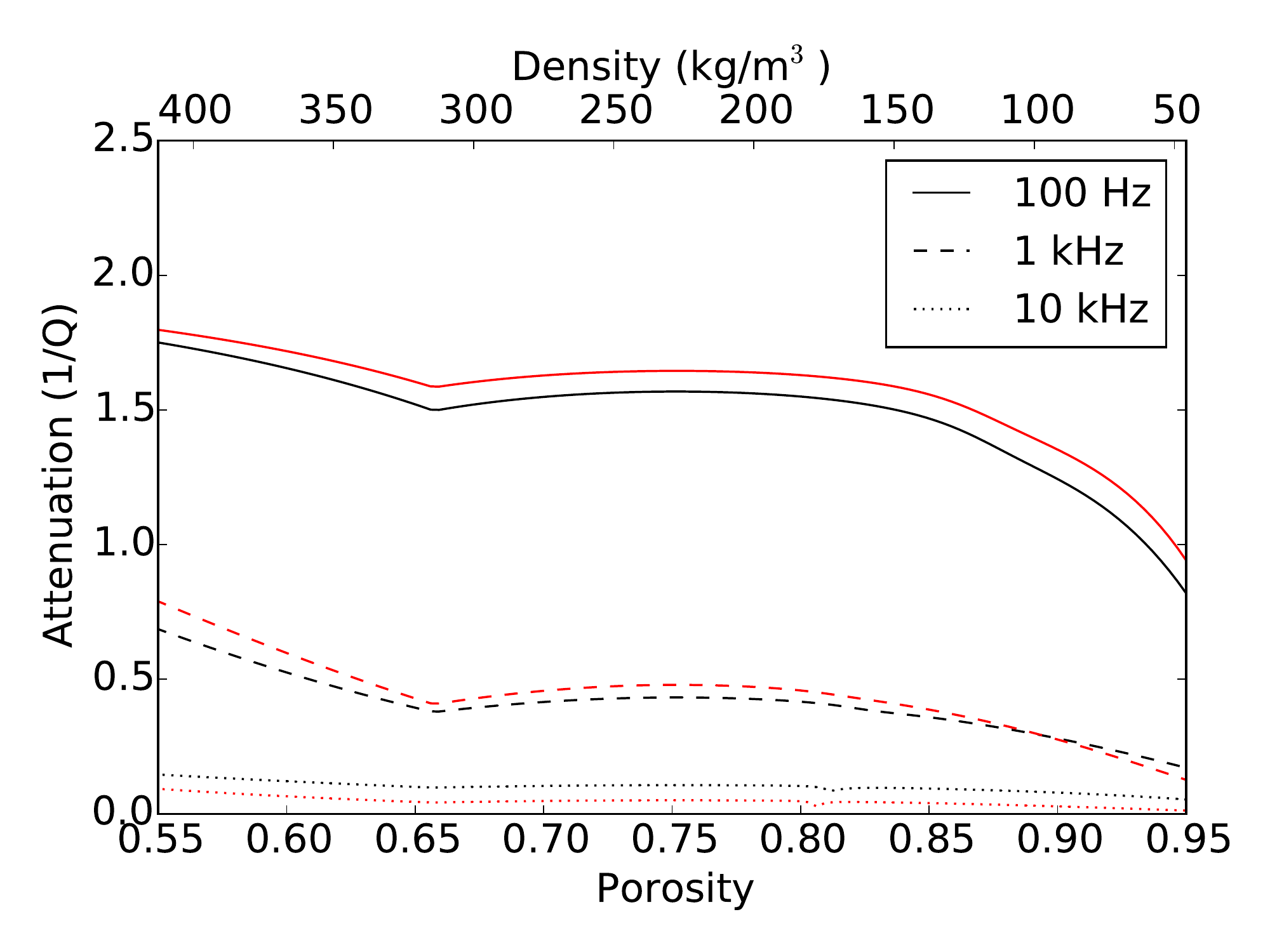}
\end{minipage} \\ 
\caption{ 
a) Phase velocity and b) attenuation for the second compressional wave for 100~Hz, 1~kHz and 10~kHz.
The black lines correspond to solutions including dynamic viscous effects considered by \citet{johnson:1987} while the red lines corresponds to solutions of Biot's \citeyearpar{biot:1956} differential equations without frequency correction.
The signs denote velocity measurements from \citet{oura:1952} and \citet{johnson:1982}.}
\label{fig:freq-dep}
\end{figure}

Phase velocity and attenuation for the second compressional wave obtained with and without using the frequency correction discussed in Section \ref{sec:viscous} are shown in Figure \ref{fig:freq-dep}.
The effect of the dynamic viscous effects are relatively small except in the range of Biot' frequency, where the phase velocity shows a moderate difference between the solutions including and neglecting a frequency correction \citep{johnson:1987}.
In contrast to the first compressional wave there is no distinctive difference in attenuation for dense and light snow for the second compressional wave.
The sharp bend in phase velocity and attenuation is due to the relationship between porosity and the specific surface area that was chosen to be constant for $\phi < 0.65$ to avoid the negative values resulting from Equation (\ref{eq:SSA}).

\begin{figure}
\centering{\includegraphics[width=80mm]{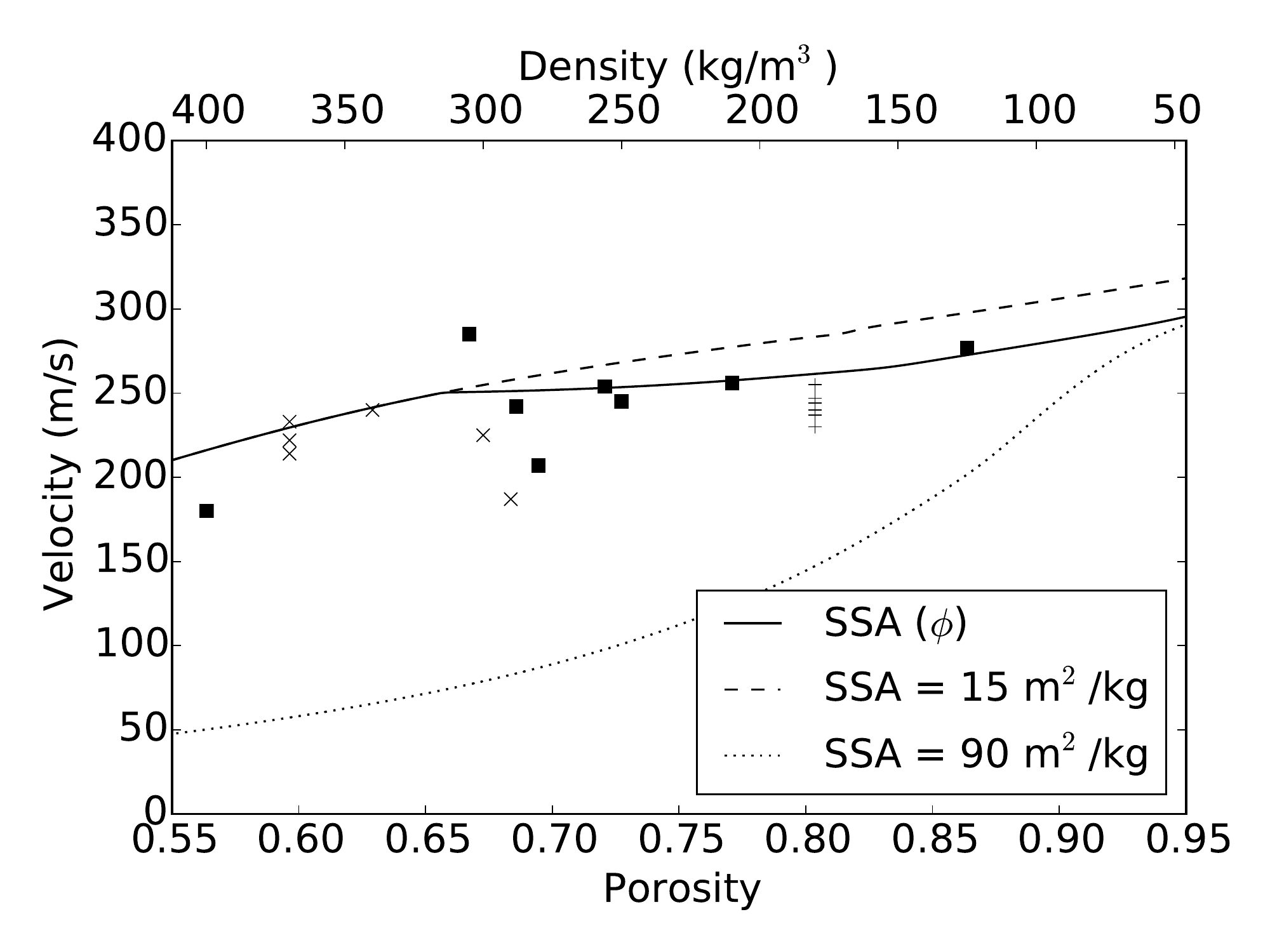}}
\caption{ 
Predicted phase velocities for the compressional wave of the second kind at 500 Hz for a specific surface area as a function of porosity (solid line) and constant values of SSA = 15 m$^2$/kg (dashed line) and SSA = 90 m$^2$/kg (dotted line).
Squares and crosses correspond to measurements from \citet{johnson:1982} and \cite{oura:1952}, respectively.}
\label{fig:v2-ssa}
\end{figure}

The variation of the phase velocity of the second compressional wave due to changes in specific surface area are shown in Figure \ref{fig:v2-ssa}. 
For fixed values of specific surface area of SSA = 15 m$^2$/kg and SSA = 90 m$^2$/kg the phase velocity at 500 Hz is plotted with dashed and dotted lines, respectively.
The solid line represents the phase velocities resulting from Equation (\ref{eq:SSA}).
The variation is larger for denser snow than for light snow where permeability is less affected by specific surface area.

\begin{figure}
\centering
\begin{minipage}[t]{3mm}
\large a) 
\end{minipage}
\begin{minipage}[t]{80mm}
\vspace{-10pt}
\includegraphics[width=1\textwidth]{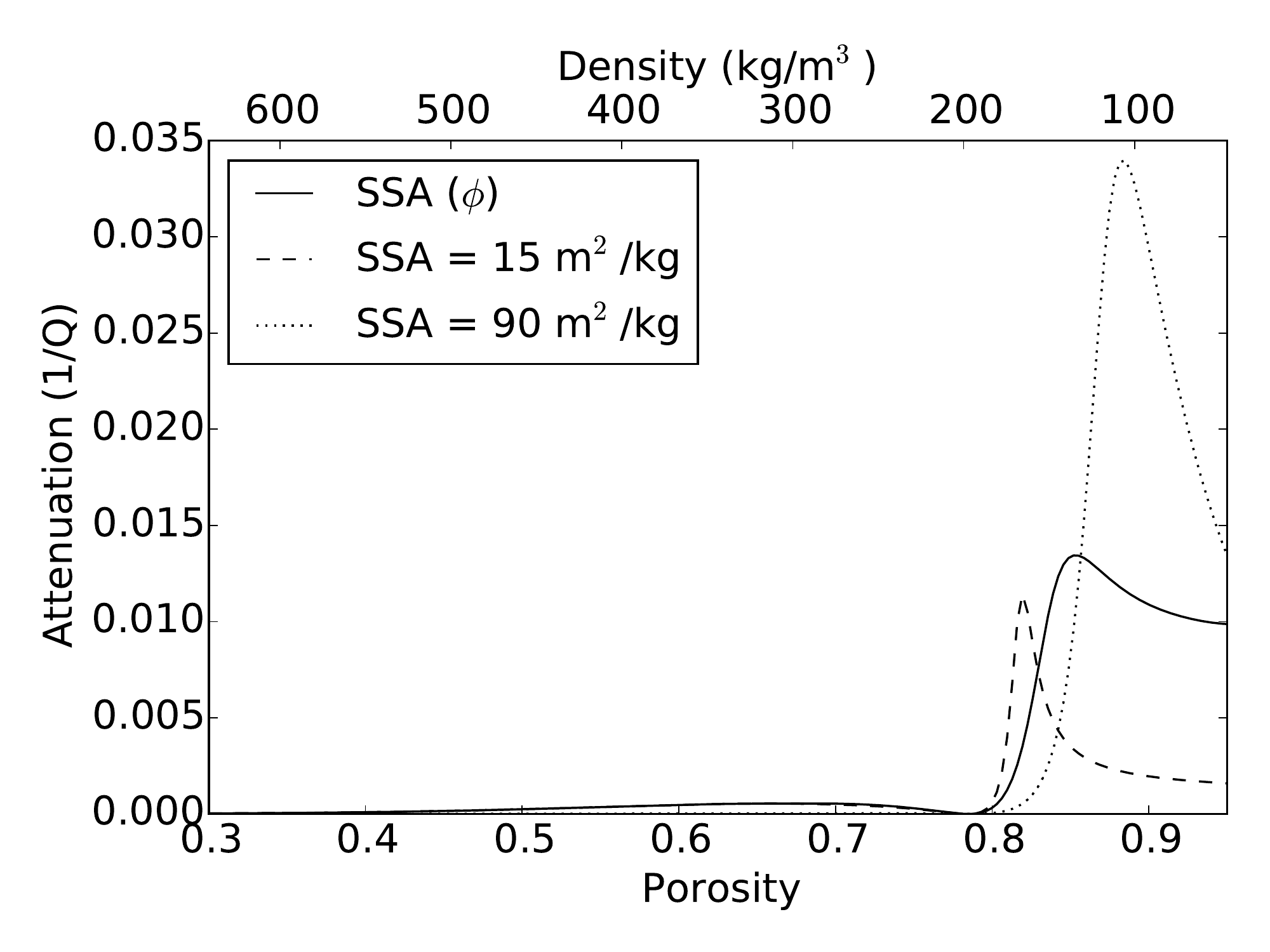}
\end{minipage}
\begin{minipage}[t]{3mm}
\large b) 
\end{minipage}
\begin{minipage}[t]{80mm}
\vspace{-10pt}
\includegraphics[width=1\textwidth]{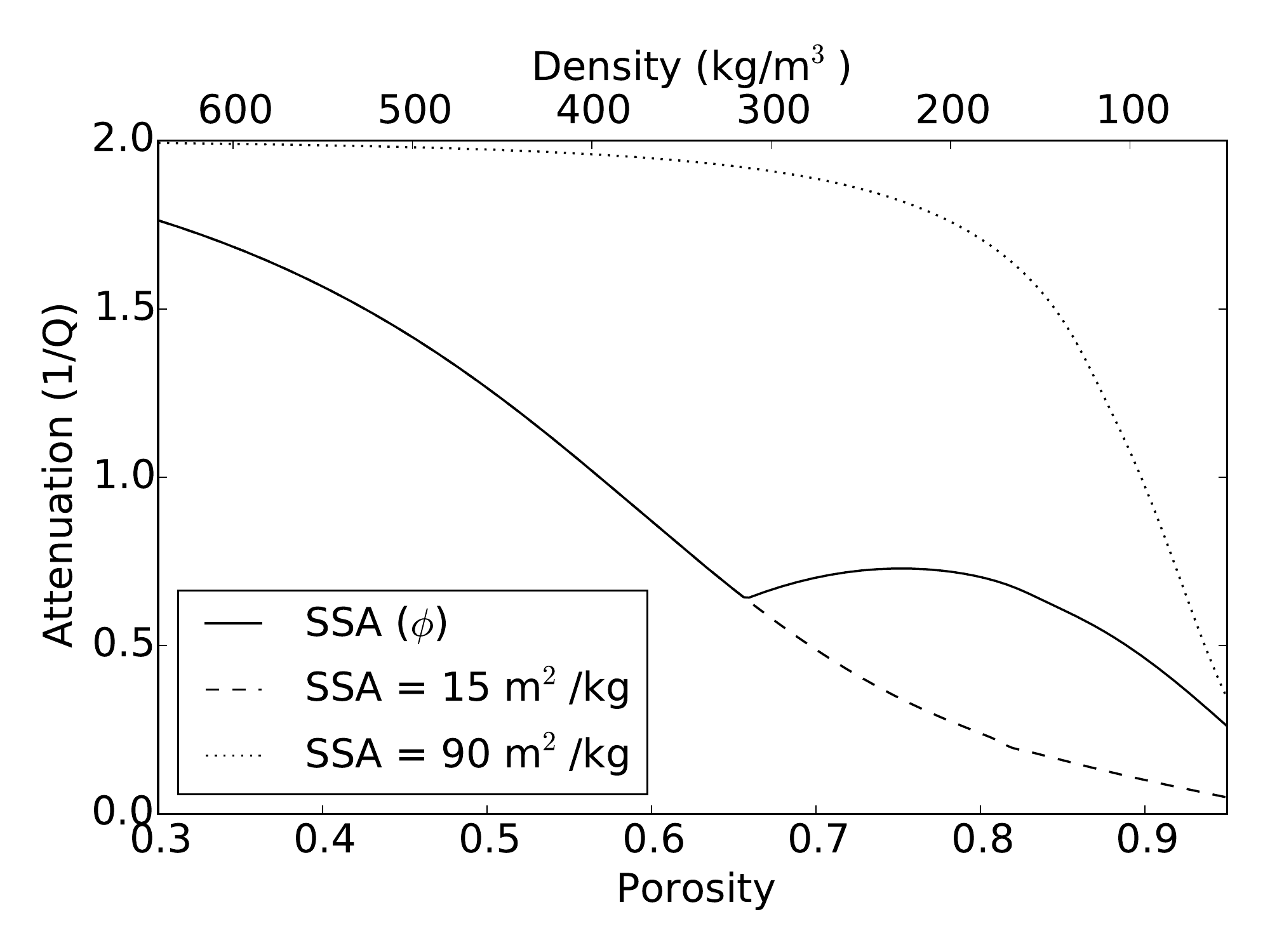}
\end{minipage} \\ 
\caption{
Predicted attenuation at 500Hz for the compressional wave of a) the first and b) second kind as a function of porosity. The dashed and dotted lines correspond to fixed values for specific surface area of SSA =  15 m$^2$/kg and SSA = 90 m$^2$/kg, respectively. The solid line corresponds to equation (\ref{eq:SSA}) and a constant value of SSA = 15 m$^2$/kg for densities above 315 kg/m$^3$. }
\label{fig:ssa-att}
\end{figure}

In Figure \ref{fig:ssa-att} attenuation for both compressional waves is shown for a constant specific surface area of SSA = 15 m$^2$/kg, and SSA = 90 m$^2$/kg, as well as for specific surface area as a function of porosity by the use of Equation \ref{eq:SSA}.
The attenuation levels of both compressional waves increase with an increase of specific surface area.
When comparing Figure \ref{fig:ssa-att}a) to Figure \ref{fig:fast-att}a) and Figure \ref{fig:ssa-att}b) to Figure \ref{fig:freq-dep}b) it becomes obvious that the effects of the specific surface area has a similar effect on attenuation as a change of frequency. 
This is an effect of scale. A higher specific surface area goes along with lower permeability and therefore has a similar effect as if the wavelength would be increased. 

\section{Discussion}

\subsection{Slow first compressional phase velocity} 

\begin{figure}
\centering{\includegraphics[width=80mm]{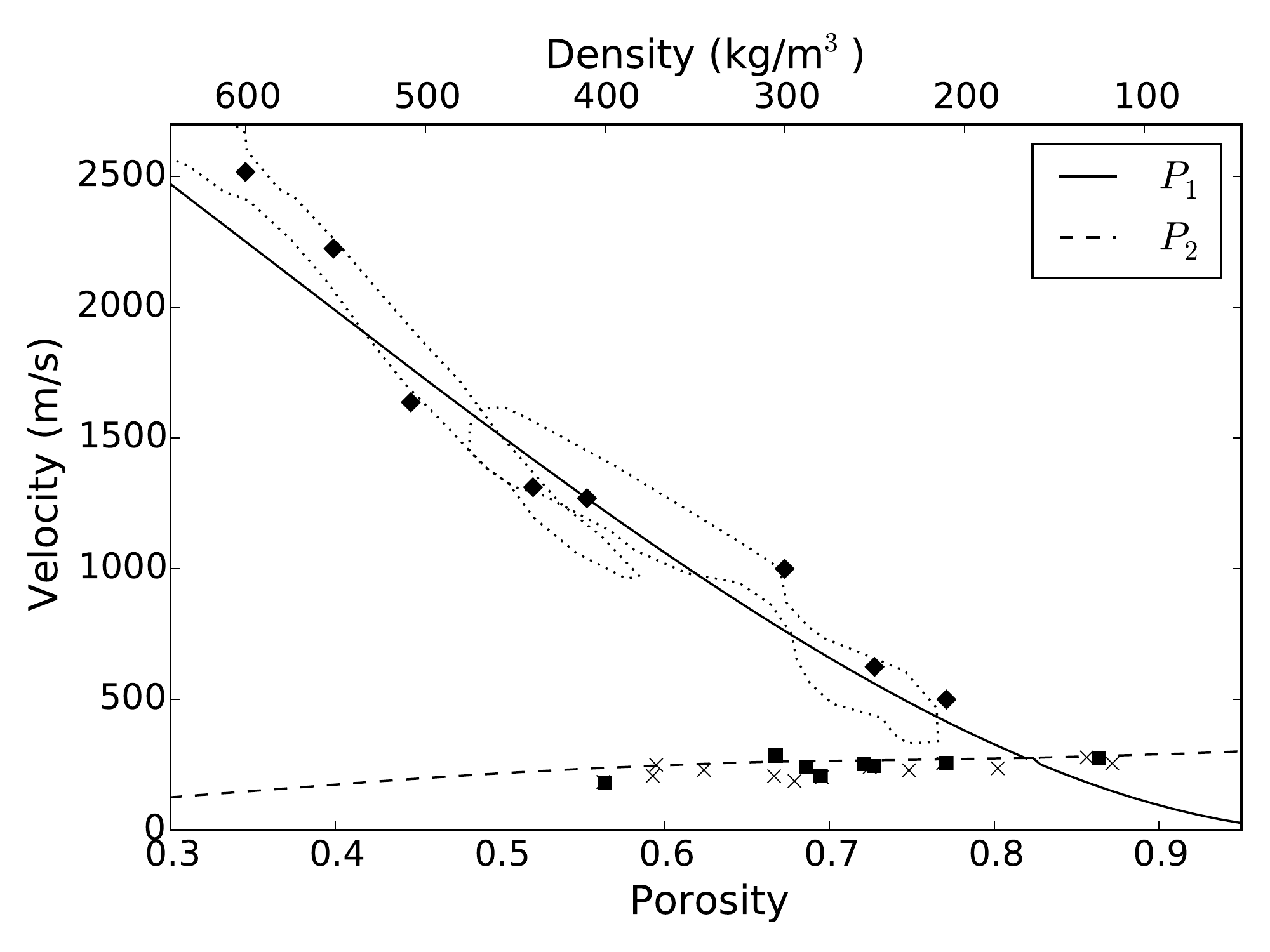}}
\caption{
Predicted velocities for compressional waves of the first (solid line) and second kind (dashed line) as a function of porosity based on empirical relationships for frame bulk and shear modulus, tortuosity and permeability. The dashed lines identify measurements of first compressional waves compiled by \citep{sommerfeld:1982} and the diamonds and squares represent wave velocity measurements compiled by \citet{johnson:1982} for compressional waves of the first and second kind, respectively.}
\label{fig:velocities}
\end{figure}

Compressional wave velocities as a function of porosity compared to measurements presented by \citet{johnson:1982} and \citet{sommerfeld:1982} are shown in Figure \ref{fig:velocities}.
The relations between porosity and the properties of the porous material, especially the strong decrease of matrix bulk modulus with increasing porosity, lead to the peculiarity that the predicted first compressional wave becomes slower than the second compressional wave for light snow with a porosity $\phi \gtrsim 0.8$.
In most materials, the second compressional wave is considerably slower than the first compressional wave and is therefore sometimes also called the `slow' wave. 
No measurements of first compressional wave with a lower phase velocity than the second compressional wave have been reported for snow.
However, a first compressional wave with lower phase velocity than the second compressional wave has been observed in high porosity reticulated foam \citep{attenborough:2012}.


From the plane wave solutions it is not immediately clear that the lower compressional velocity in light snow corresponds to the first compressional wave mode as there are no explicit rules to choose the signs of the square roots.
To illustrate that it is indeed the velocity of the first compressional wave that is slower than the phase velocity of the second compressional wave, two numerical simulations of wave propagation in poroelastic materials were performed. In the first simulation the homogeneous poroelastic material corresponds to snow with a porosity $\phi = 0.7$, where the first compressional wave is expected to be faster than the second compressional wave. 
In the second simulation the porosity of the snow is chosen to be $\phi = 0.9$ and the first compressional wave is expected to be slower than the second compressional wave. 
The relations from Section \ref{sec:snow-model} are used to characterize the remaining porous material properties.

For the simulation a pseudo spectral modeling code was used \citep{sidler:2010a}. To avoid differences due to the source characteristics a pressure source with a waveform of a Ricker wavelet with a central frequency of 500~Hz is placed in the air above the poroelastic material and the boundary conditions are assumed to be of the 'open pore' type \citep{deresiewicz:1963}.
The field variables of the simulation are the velocity of the solid frame, the relative velocity of the pore fluid to the solid frame, the stress tensor, and the pore pressure.
Please note that these are particle velocities and should not be confused with the phase velocities discussed before.
Even though the wave modes do not travel independently of each other in the numerical simulation, in a homogeneous material, the motion of the solid frame corresponds roughly to the first compressional wave that travels mainly in the solid frame.
Likewise, the relative motion of the pore fluid corresponds roughly to the second compressional wave (see also equations (\ref{eq:fast-wave}) and (\ref{eq:slow-wave})). 

\begin{figure*}
\centering
\large a) 
\begin{minipage}[t]{75mm}
\vspace{-10pt}
\includegraphics[width=1\textwidth]{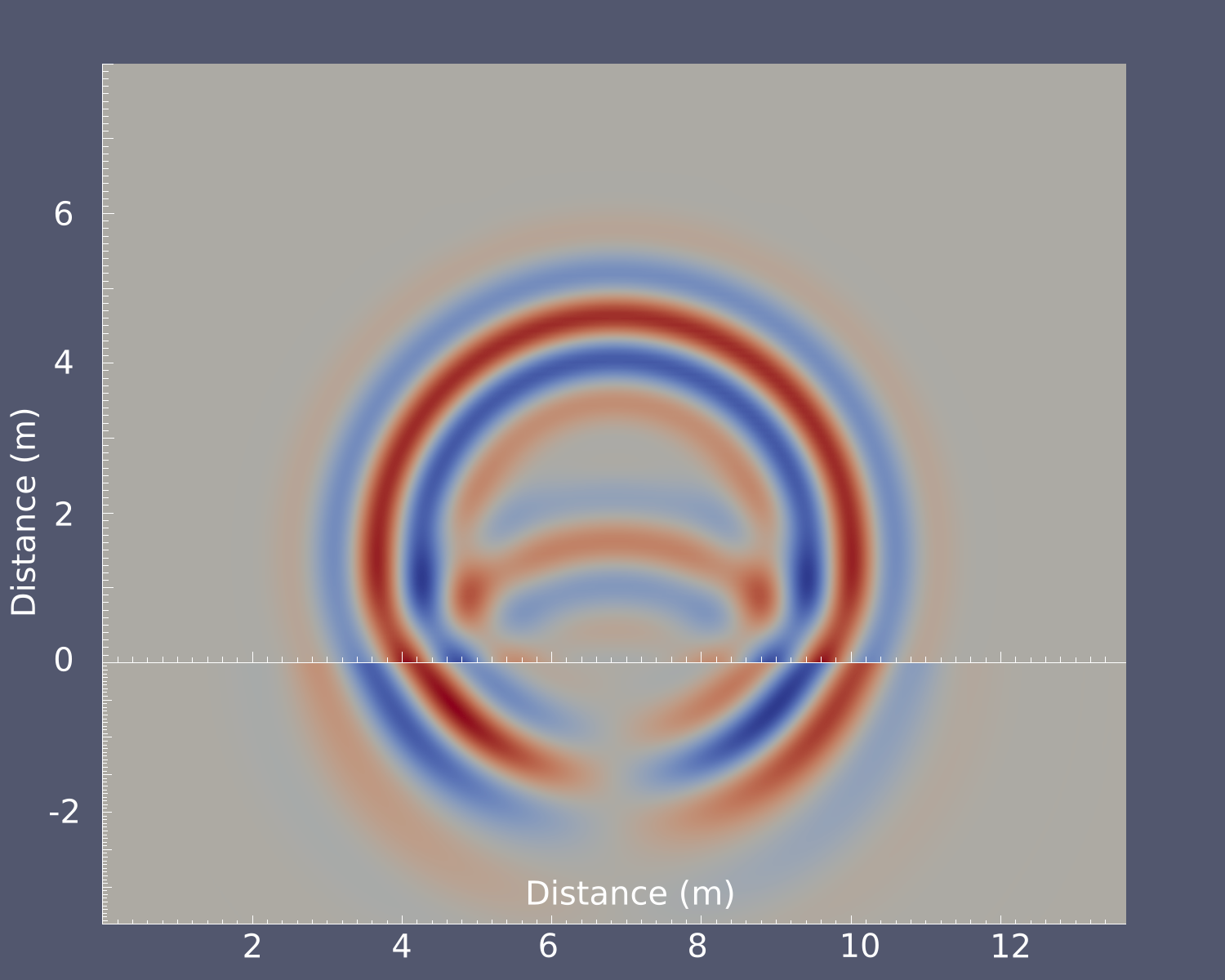}
\end{minipage}
\begin{minipage}[t]{2mm}
\large b) 
\end{minipage}
\begin{minipage}[t]{75mm}
\vspace{-10pt}
\includegraphics[width=1\textwidth]{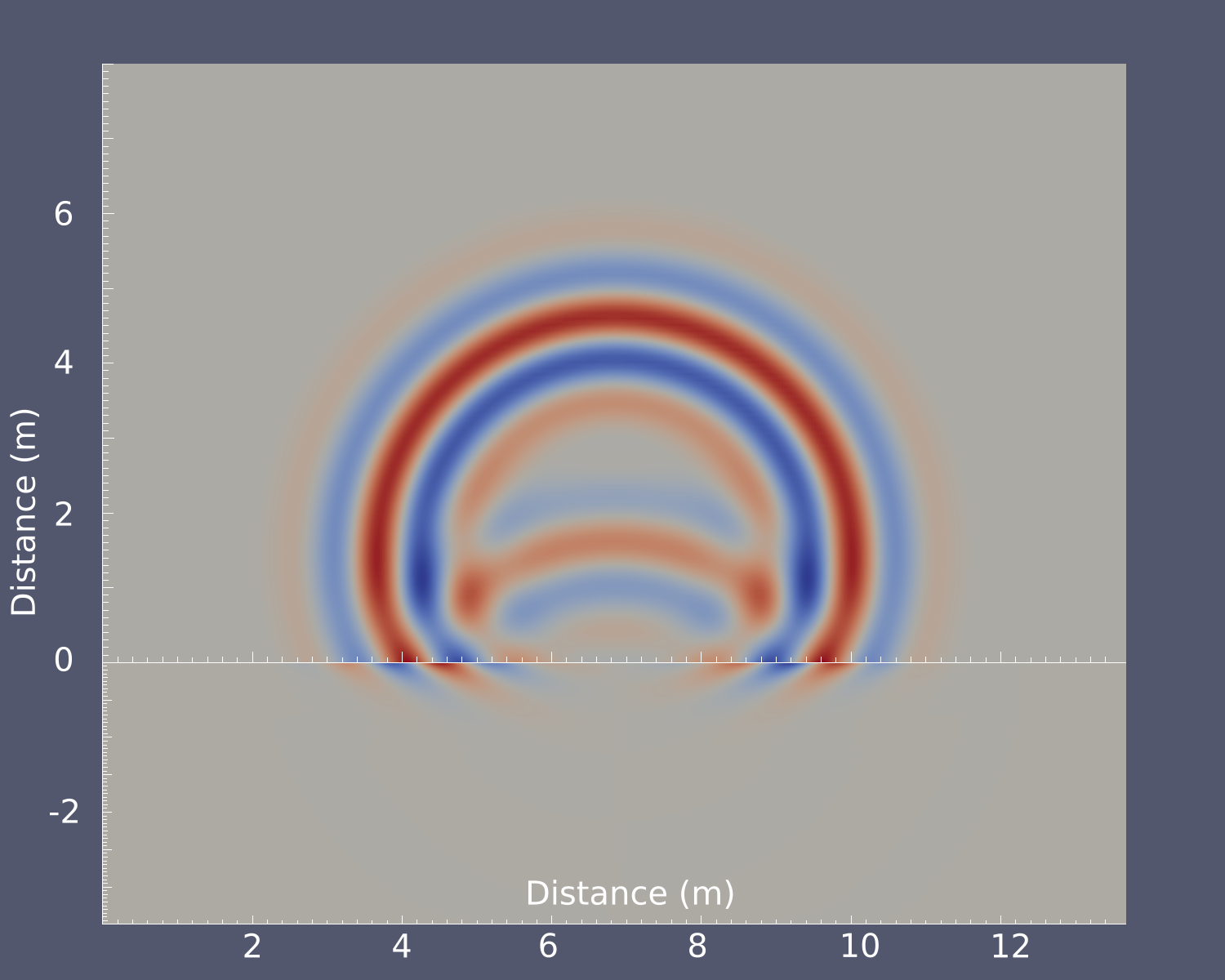}
\end{minipage} \\ 
\large c) 
\begin{minipage}[t]{75mm}
\vspace{-10pt}
\includegraphics[width=1\textwidth]{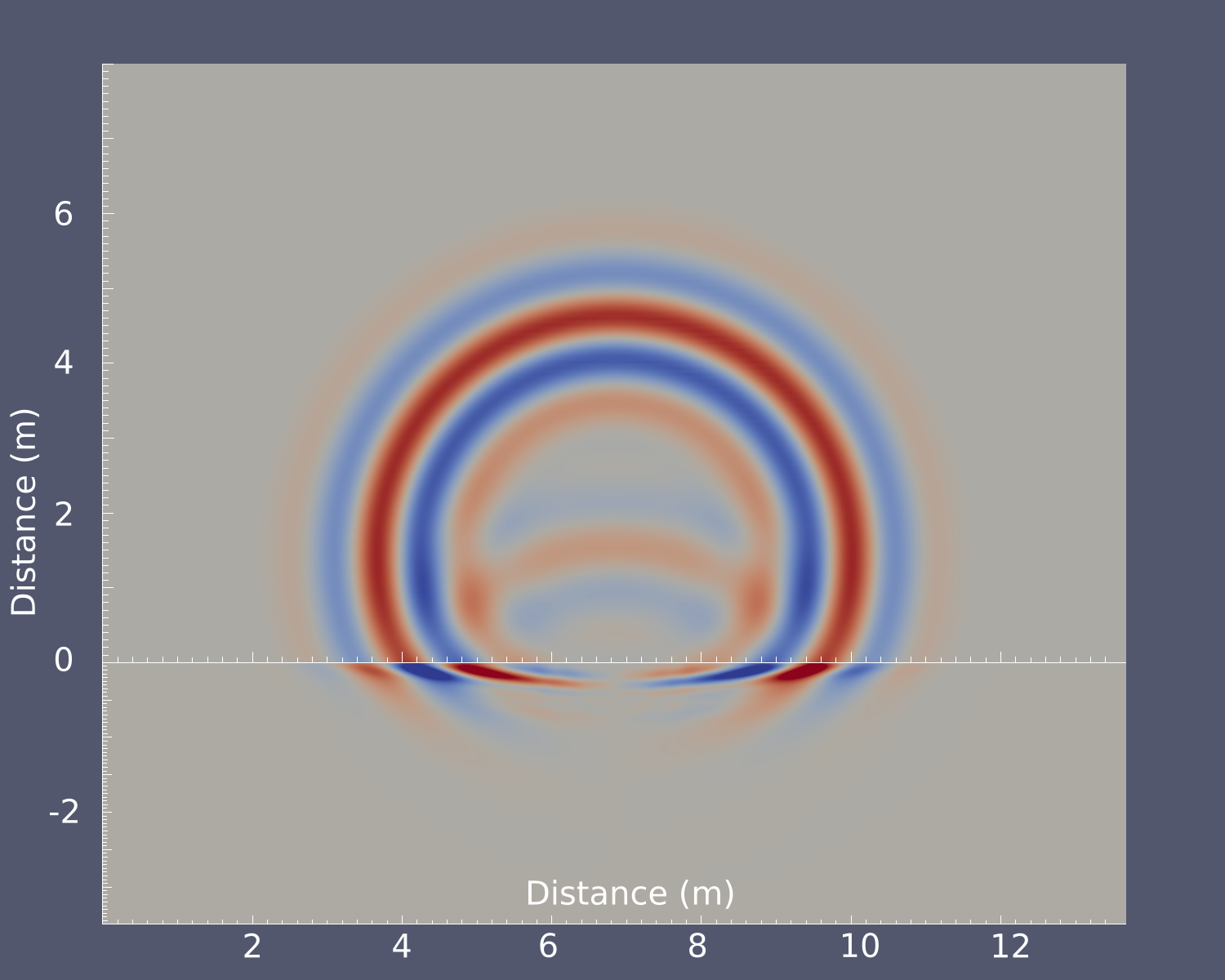}
\end{minipage}
\begin{minipage}[t]{2mm}
\large d) 
\end{minipage}
\begin{minipage}[t]{75mm}
\vspace{-10pt}
\includegraphics[width=1\textwidth]{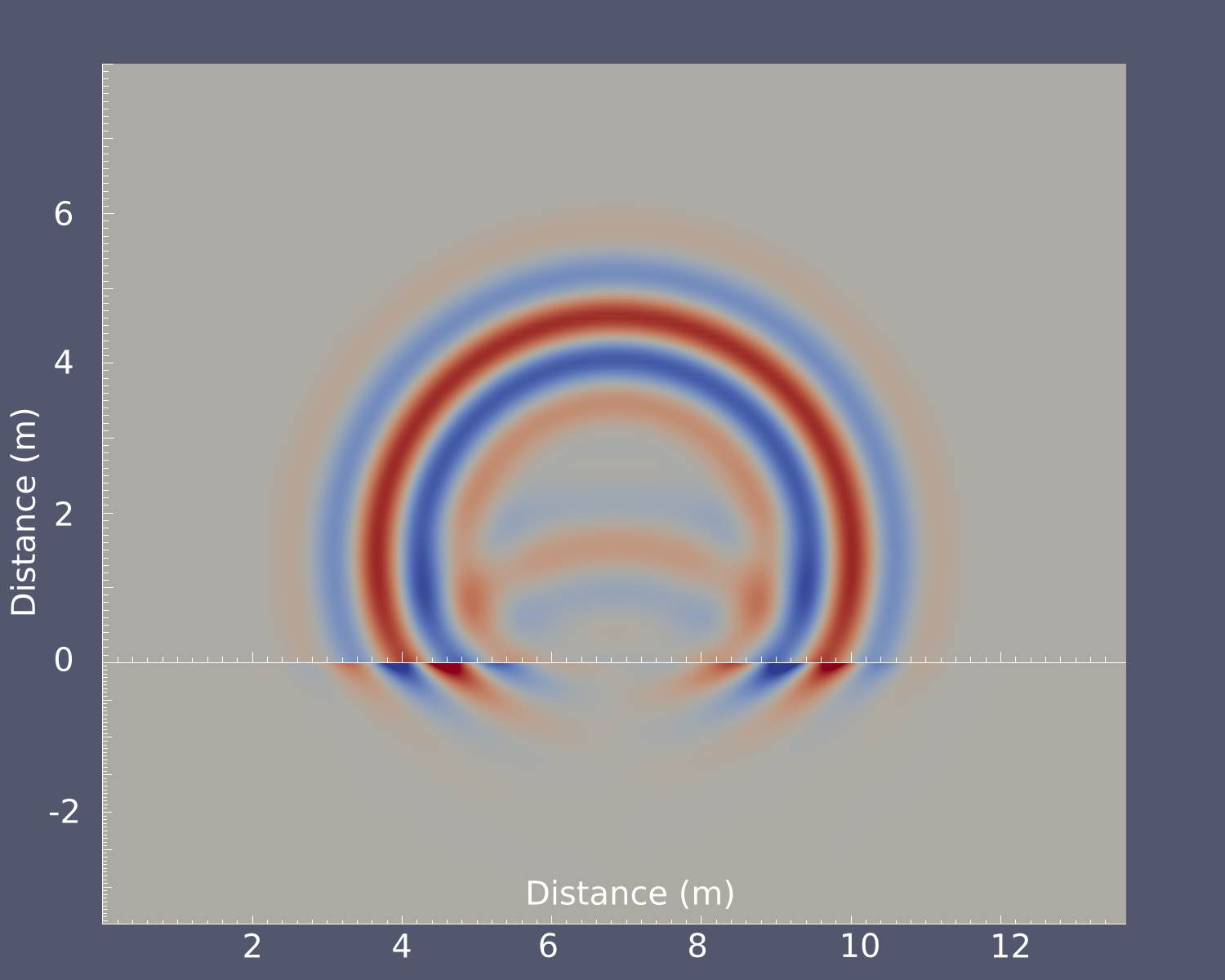}
\end{minipage} \\ 
\caption{\label{fig:fast-slow-wave} Snapshots after 15.6~ms of a numerical simulation of a pressure source in the air over snowpacks with a porosity ( a) and b) ) $\phi = 0.7$ and ( c) and d) ) $\phi = 0.9$. The horizontal components of the (a) and c)) velocity of the porous frame and (b) and d)) the relative velocity of the pore fluid to the porous frame are shown. It can be seen that in the highly porous material ($\phi = 0.9$), the first compressional wave ( c) ) is slower than the second compressional wave ( d) ).}
\end{figure*}

In Figure \ref{fig:fast-slow-wave} snapshots of horizontal components of the velocity of the porous frame and the relative velocities of the pore fluid to the porous frame are shown 15.6~ms after triggering the acoustic source for the two simulations.
The velocity of the second compressional wave is almost the same in both simulations (Figure \ref{fig:fast-slow-wave} b) and d) ).
However, it is evident that the first compressional wave is faster than the second compressional wave in the first simulation (Figure \ref{fig:fast-slow-wave} a) ) and slower than the second compressional wave in the second simulation (Figure \ref{fig:fast-slow-wave} c) ).

\subsection{Increased sound absorption of light snow} 

The attenuation levels of the first compressional wave differs significantly for light snow with a porosity $\phi \gtrsim 0.8$ and snow with a lower porosity.
This separation corresponds roughly to a separation between freshly fallen and aged snow \citep{judson:2000}.
In between the two porosity ranges the attenuation vanishes completely as the two wave modes have the same velocity and the viscous effects leading to attenuation are not effective.
The sound absorption above ground is a complex combination of effects involving amongst others the interference of incident and reflected waves, the reflection coefficient, geometrical spreading as well as surface and non geometrical waves \citep{embleton:1996}.
However, it is clear that if the reflection coefficient of the ground decreased also the sound level above the ground decreases \citep{watson:1948,nicolas:1985}.
Due to the high porosity of snow and the open pore boundary conditions, the pressure of the air above the snow pack interacts mainly with the air in the pore space and little energy is transmitted into the ice frame. As the velocity of the second compressional wave is almost equal to the velocity of the air above the the snow, there is almost no impedance contrast that would lead to a reflection.
The low velocities of the first compressional wave in snow with porosity $\phi > 0.8$ and the corresponding higher attenuation further decrease the impedance contrast and also reduce the contribution of refracted waves.

\section{Conclusions}

A method to predict phase velocities and plane wave attenuation of acoustic waves as a function of snow porosity is presented. The method is based on Biot's \citeyearpar{biot:1956} model of wave propagation in porous materials and uses empirical relationships to assess tortuosity, permeability, bulk, and shear moduli as a function of porosity. The properties of the ice frame of the snow and air as the pore fluid are assumed to be constant.
The method is not restricted to porosity as a single degree of freedom and additional information on specific surface area or any other of the properties characterizing a Biot-type porous medium can be readily incorporated.

For light snow with a porosity $\phi \gtrsim 0.8$ the particularity is found that the velocities of the compressional wave of the first kind is slower than the phase velocities of the compressional wave of the second kind which is commonly referred to as the `slow' wave. Such a reversal of the velocities of the compressional waves has been observed in reticulated foam before and is due to the week structure of the ice matrix in fresh and light snow. 
The wave velocity reversal is a relatively sharp boundary for the attenuation level of the first compressional wave which is orders of magnitudes larger for highly porous snow. 
This finding is in accordance with the well known observation that freshly fallen snow absorbs most of the ambient noise, while after a relatively short time this absorbing behavior vanishes.

The first compressional wave is sensitive mainly to matrix and shear bulk modulus. 
A variation of $\sim$25 \% in both, shear and matrix bulk modulus can characterize the variability in the measured velocities. 
The attenuation of the second compressional wave decreases with increasing porosity and is considerably higher than for the first compressional wave. Also frequency dependence of the attenuation is considerably more distinct than for the first compressional wave. The velocity of the second compressional wave depends strongly on tortuosity, permeability, and the related specific surface area.
The variation of measured wave velocities for the second compressional wave can be obtained by altering the tortuosity by $\sim$30~\% or by altering the permeability by $\sim$50~\%.

This method is a viable prerequisite for numerical modeling of acoustic wave propagation in snow, which allows, for example, to assess the design of acoustic experiments to probe for snow properties or to assess the role of acoustic wave propagation in artificial or skier triggered snow avalanche releases.

Further research will address the presence of liquid water in the pore space, a more complete analysis of sound absorption above snow of different porosity, and numerical simulations of explosive avalanche mitigation experiments.

%
%

\section{Acknowledgements}
This research was founded by a fellowship of the Swiss National Science Foundation.

%
%

%
%

\bibliography{0-references}
\bibliographystyle{igs}

%
%
\end{document}